\documentclass[a4paper,12pt]{article}
\usepackage{amssymb, amsmath}
\usepackage[curve]{xypic}

\providecommand{\abs}[1]{\lvert#1\rvert}

\providecommand{\Spin}{\textnormal{Spin}}
\providecommand{\SO}{\textnormal{SO}}

\providecommand{\CCl}{\mathbb{C}\textnormal{l}}

\providecommand{\id}{\textnormal{id}}

\providecommand{\Ker}{\textnormal{Ker}}
\providecommand{\IIm}{\textnormal{Im}}

\providecommand{\Hom}{\textnormal{Hom}}
\providecommand{\Ker}{\textnormal{Ker}}

\providecommand{\rk}{\textnormal{rk}}
\providecommand{\Int}{\textnormal{Int}}
\providecommand{\ch}{\textnormal{ch}}
\providecommand{\PD}{\textnormal{PD}}

\providecommand{\ev}{\textnormal{ev}}
\providecommand{\odd}{\textnormal{odd}}

\begin{document}

\begin{titlepage}
\titlepage
\rightline{SISSA 31/2008/FM-EP}
\vskip 2.5cm
\centerline{{ \bf \LARGE Comparing two approaches to the K-theory}}
\vskip 0.5cm
\centerline{{ \bf \LARGE classification of D-branes}}
\vskip 1.7truecm

\begin{center}
{\bf \large Fabio Ferrari Ruffino and Raffaele Savelli}
\vskip 1.5cm
\em 
International School for Advanced Studies (SISSA/ISAS) \\ 
Via Beirut 2, I-34014, Trieste, Italy\\
and Istituto Nazionale di Fisica Nucleare (INFN), sezione di Trieste

\vskip 2.5cm

\bf Abstract
\end{center}

\normalsize We consider the two main classification methods of D-brane charges via K-theory, in type II superstring theory with vanishing $B$-field: the Gysin map approach and the one based on the Atiyah-Hirzebruch spectral sequence. Then, we find out an explicit link between these two approaches: the Gysin map provides a representative element of the equivalence class obtained via the spectral sequence. We also briefly discuss the case of rational coefficients, characterized by a complete equivalence between the two classification methods.

\vskip2cm

\vskip1.5\baselineskip

\vfill
 \hrule width 5.cm
\vskip 2.mm
{\small 
\noindent }
\begin{flushleft}
ferrariruffino@gmail.com, savelli@sissa.it
\end{flushleft}
\end{titlepage}

\tableofcontents

\newtheorem{Theorem}{Theorem}[section]
\newtheorem{Lemma}[Theorem]{Lemma}
\newtheorem{Corollary}[Theorem]{Corollary}
\newtheorem{Rmk}[Theorem]{Remark}
\newtheorem{Def}{Definition}[section]
\newtheorem{ThmDef}[Theorem]{Theorem - Defintion}

\vspace{1cm}

\section{Introduction}

K-theory provides a good tool to classify D-brane charges in type II superstring theory \cite{Evslin,OS}. In the case of vanishing $B$-field, there are two main approaches in the literature. The first one consists of applying the Gysin map to the gauge bundle of the D-brane, obtaining a K-theory class in space-time \cite{MM}. This approach is motivated by the Sen conjecture, stating that a generic configuration of branes and antibranes with gauge bundle is equivalent, via tachyon condensation, to a stack of coincident space-filling brane-antibrane pairs equipped with an appropriate K-theory class \cite{Sen}. The second approach consists of applying the Atiyah-Hirzebruch spectral sequence (AHSS, \cite{AH}) to the Poincar\'e dual of the homology class of the D-brane: such a sequence rules out some cycles affected by global world-sheet anomalies, e.g.\ Freed-Witten anomaly \cite{FW}, and quotients out some cycles which are actually unstable, e.g.\ MMS-instantons \cite{MMS}. We assume for simplicity that the space-time and the D-brane world-volumes are compact. For a given filtration of the space-time $S = S^{10} \supset S^{9} \supset \cdots \supset S^{0}$, the second step of AHSS is the cohomology of $S$, i.e.\ $E^{p,\,0}_{2}(S) \simeq H^{p}(S, \mathbb{Z})$, while the last step of AHSS is given by (up to canonical isomorphism):
	\[E^{p,\,0}_{\infty}(S) \simeq \frac{\, \Ker \bigl( K^{p}(S) \longrightarrow K^{p}(S^{p-1}) \bigr) \,} {\Ker \bigl( K^{p}(S) \longrightarrow K^{p}(S^{p}) \bigr)} \; .
\]
Hence, given a D-brane world-volume $WY_{p}$ of codimension $10 - (p+1) = 9-p$, with gauge bundle $E \rightarrow WY_{p}$ of rank $q$, if the Poincar\'e dual of $WY_{p}$ in $S$ survives until the last step of AHSS, it determines a class $\{\PD_{S}[q \cdot WY_{p}]\} \in E^{9-p,\,0}_{\infty}(S)$ whose representatives belong to $\Ker (K^{9-p}(S) \longrightarrow K^{9-p}(S^{8-p}))$.

These two approaches give different information, in particular AHSS does not take into account the gauge bundle: the aim of the present work is to relate them. We briefly anticipate the result. For a D$p$-brane with world-volume $WY_{p} \subset S$ and gauge bundle $E \rightarrow WY_{p}$ of rank $q$, let $i: WY_{p} \rightarrow S$ be the embedding and $i_{!}: K(WY_{p}) \rightarrow K^{9-p}(S)$ the Gysin map. We will show that $i_{!}(E) \in \Ker (K^{9-p}(S) \rightarrow K^{9-p}(S^{8-p}))$ and that:
	\[\{\PD_{S}[q \cdot WY_{p}]\}_{E^{9-p, \,0}_{\infty}} = [i_{!}(E)] \; .
\]
Thus, we must first use AHSS to detect possible anomalies, then we can use the Gysin map to get the charge of a non-anomalous brane: such a charge belongs to the equivalence class reached by AHSS, so that the Gysin map gives more detailed information. For further remarks about this we refer to the conclusions.

Moreover, we compare this picture with the case of rational coefficients. It is known that the Chern character provides isomorphisms $K(S) \otimes_{\mathbb{Z}} \mathbb{Q} \simeq H^{\ev}(S, \mathbb{Q})$ and $K^{1}(S) \otimes_{\mathbb{Z}} \mathbb{Q} \simeq H^{\odd}(S, \mathbb{Q})$, and that AHSS with rational coefficients degenerates at the second step, i.e.\ at the level of cohomology. Therefore, we gain a complete equivalence between the two K-theoretical approaches, being both equivalent to the old cohomological classification.

The paper is organized as follows. In section 2 we discuss in detail the physical context underlying the K-theory classification of D-branes. In section 3 and 4 we introduce the topological tools needed to formulate our result, which is stated and proven in section 5. In section 6 we draw our conclusions.

\section{Physical motivations}

For simplicity we assume the ten-dimensional space-time $S$ to be a compact manifold, so that also the D-brane world-volumes are compact. This seems not physically reasonable, but it has more meaning if we suppose to have performed the Wick rotation in space-time, so that we work in a euclidean setting. In this setting we loose the physical interpretation of the D-brane world-volume as a volume moving in time and of the charge $q$ (actually all the homology class $[q \cdot Y_{p,t}]$ for $Y_{p,t}$ the restriction of the world-volume at an instant $t$ in a fixed reference frame) as a charge conserved in time. Thus, rather than considering the homology class of the D-brane volume at every instant of time, we prefer to consider the homology class of the entire world-volume in $S$, using standard homology with compact support.

\subsection{Classification}

For a D$p$-brane with $(p+1)$-dimensional world-volume $WY_{p}$ and charge $q$ we consider the corresponding homology class in $S$:
\begin{equation}\label{HomologicalClassification}
	[q \cdot WY_{p}] \in H_{p+1}(S, \mathbb{Z}) = \frac{Z_{p+1}(S, \mathbb{Z})}{B_{p+1}(S, \mathbb{Z})} = \mathbb{Z}^{b_{p+1}} \oplus_{i} \mathbb{Z}_{p_{i}^{n_{i}}}
\end{equation}
where $Z_{p+1}(S, \mathbb{Z})$ denotes the group of singular $(p+1)$-cycles of $S$, $B_{p+1}(S, \mathbb{Z})$ the subgroup of $(p+1)$-boundaries, $b_{p+1}$ the $(p+1)$-th Betti number of $S$, and $p_{i}$ is a prime number for every $i$. For what will follow, it is convenient to consider the cohomology of $S$ rather than the homology. Hence, denoting by $\PD_{S}$ the Poincar\'e duality map on $S$,\footnote{As we said above, we are assuming for simplicity that the space-time is a compact manifold (without singularities), and we also suppose it is orientable, thus Poincar\'e duality holds.} we define the \emph{charge density}:
\begin{equation}\label{PDS}
	\PD_{S}[q \cdot WY_{p}] \in H^{9-p}(S, \mathbb{Z}) = \frac{Z^{9-p}(S, \mathbb{Z})}{B^{9-p}(S, \mathbb{Z})} = \mathbb{Z}^{b_{p+1}} \oplus_{i} \mathbb{Z}_{p_{i}^{n_{i}}}
\end{equation}
where $Z^{9-p}(S, \mathbb{Z})$ is the group of singular $(9-p)$-cocyles and $B^{9-p}(S, \mathbb{Z})$ the subgroup of $(9-p)$-coboundaries. This classification encounters some problems due to the presence of quantum anomalies. Two remarkable examples are the following:
\begin{itemize}
	\item a brane wrapping a cycle $WY_{p} \subset S$ is Freed-Witten anomalous if its third integral Stiefel-Whitney class $W_{3}(WY_{p})$ is not zero, hence not all the cycles are allowed \cite{FW, Evslin};
	\item given a world-volume $WY_{p}$ with $W_{3}(WY_{p}) \neq 0$, it can be interpreted as an MMS-instanton in the minkowskian setting \cite{MMS, Evslin}; in this case there are cycles intersecting $WY_{p}$ in $\PD_{WY_{p}}(W_{3}(WY_{p}))$ which, although homologically non-trivial in general, are actually unstable.
\end{itemize}
The two points above imply that:
\begin{itemize}
	\item the numerator $Z_{p+1}(S, \mathbb{Z})$ of \eqref{HomologicalClassification} is too large, since it contains anomalous cycles;
	\item the denominator $B_{p+1}(S, \mathbb{Z})$ of \eqref{HomologicalClassification} is too small, since it does not cut all the unstable charges.
\end{itemize}
There are other possible anomalies, although not yet completely understood, some of which are probably related to homology classes not representable by a smooth submanifold \cite{ES, BHK, Evslin}.

\paragraph{}We start by considering the case of world-volumes of \emph{even} codimension in $S$, i.e.\ we start with IIB superstring theory. To solve the problems mentioned above, one possible tool seems to be the Atiyah-Hirzebruch spectral sequence \cite{AH}. Choosing a finite simplicial decomposition \cite{Hatcher} of the space-time manifold $S$, and considering the filtration $S = S^{10} \supset \cdots \supset S^{0}$ for $S^{i}$ the $i$-th dimensional skeleton, such a spectral sequence starts from the even-dimensional simplicial cochains of $S$ and, after a finite number of steps, it stabilizes to the graded group $E_{\infty}^{\ev, \,0}(S) = \bigoplus_{2k}K_{2k}(S) / K_{2k+1}(S)$. Here $K_{q}(S)$ is the kernel of the natural restriction map from the K-theory group of $S$, which we denote by $K(S)$, to the K-theory group of $S^{q-1}$, which we call $K(S^{q-1})$: i.e.\ $K_{q}(S) = \Ker(K(S) \rightarrow K(S^{q-1}))$. We also use the notaion $E_{\infty}^{2k, \,0}(S) = K_{2k}(S) / K_{2k+1}(S)$, so that $E_{\infty}^{\ev, \,0}(S) = \bigoplus_{2k} E_{\infty}^{2k, \,0}(S)$. We can start from a representative of the Poincar\'e dual of the D-brane $\PD_{S}[q \cdot WY_{p}]$, which in our hypotheses is even-dimensional, and, if it survives until the last step, we arrive at a class $\{\PD_{S}[q \cdot WY_{p}]\} \in K_{9-p}(S) / K_{9-p+1}(S)$. The even boundaries $d_{2}, d_{4}, \ldots$ of this sequence are 0, hence the important ones are the odd boundaries. In particular, one can prove that:
\begin{itemize}
	\item $d_{1}$ coincides with the ordinary coboundary operator, hence the second step is the even cohomology of $S$ \cite{Segal, AH};
	\item the cocycles not living in the kernel of $d_{3}$ are Freed-Witten anomalous, while the cocycles contained in its image are unstable because of the presence of MMS-instantons \cite{Evslin, MMS}.
\end{itemize}
As we will say in a while, there are good reasons to use K-theory to classify D-brane charges, hence, although the physical meaning of higher order boundaries is not completely clear, the behaviour of $d_{3}$ and the fact that the last step is directly related to K-theory suggest that the class $\{\PD_{S}[q \cdot WY_{p}])\} \in E_{\infty}^{9-p,0}(S)$ is a good candidate to be considered as the charge of the D-brane. Summarizing, we saw two ways to classify D-brane cycles and charges:
\begin{itemize}
	\item the homological classification, i.e.\ $[q \cdot WY_{p}] \in H_{p+1}(S, \mathbb{Z})$;
	\item the classification via the Atiyah-Hirzebruch spectral sequence, i.e.\ $\{\PD_{S}[q \cdot WY_{p}]\} \in E_{\infty}^{9-p, \,0}(S)$.
\end{itemize}

\subsection{K-theory from the Sen conjecture}

\subsubsection{Gauge and gravitational couplings}

Up to now we have only considered the cycle wrapped by the D-brane world-volume. There are other important features: the gauge bundle and the embedding in space-time, which enter in the action via the two following couplings:
\begin{itemize}
	\item the gauge coupling through the Chern character \cite{LM} of the Chan-Paton bundle;
	\item the gravitational coupling through the $\hat{A}$-genus \cite{LM} of the tangent and the normal bundle of the world-volume.
\end{itemize}
The unique non-anomalous form of these couplings, computed by Minasian and Moore in \cite{MM}, is:
\begin{equation}\label{action}
	S = \int_{WY_{p}} \textstyle i^{*}C \wedge \ch(E) \wedge e^{\frac{d}{2}} \wedge \frac{\,\sqrt{\hat{A}(T(WY_{p}))}\,} {\sqrt{\hat{A}(N(WY_{p}))}}
\end{equation}
where $i: WY_{p} \rightarrow S$ is the embedding, $E$ is the Chan-Paton bundle, $T(WY_{p})$ and $N(WY_{p})$ are the tangent bundle and the normal bundle of $WY_{p}$ in $S$, and $d \in H^{2}(WY_{p}, \mathbb{Z})$ is a class whose restriction mod 2 is the second Stiefel-Whitney class of the normal bundle $w_{2}(N(WY_{p}))$. The polyform that multiplies $i^{*}C$ has 0-term equal to $\ch_{0}(E) = \rk(E)$, hence \eqref{action} is an extension of the usual minimal coupling $q\int_{WY_{p}} i^{*}C_{p+1}$ for $q = \rk(E)$: the charge of the D-brane coincides with the rank of the gauge bundle (up to a normalization constant). In the case of anti-branes, we have to allow for negative charges, hence the gauge bundle is actually a \emph{K-theory class}: a generic class $E - F$ can be interpreted as a stack of pairs of a brane $Y$ and an anti-brane $\overline{Y}$ with gauge bundle $E$ and $F$ respectively. For $i_{\#}: H^{*}(WY_{p}, \mathbb{Q}) \rightarrow H^{*}(S, \mathbb{Q})$ the Gysin map in cohomology \cite{Hirzebruch, OS}, we define the \emph{charge density}:
\begin{equation}\label{QWY}
	Q_{WY_{p}} = i_{\#} \Bigl( \ch(E) \wedge e^{\frac{d}{2}} \wedge \textstyle\frac{\,\sqrt{\hat{A}(T(WY_{p})})\,} {\sqrt{\hat{A}(N(WY_{p}))}} \Bigr) \; .
\end{equation}
Since new terms have appeared in the charge, we should discuss also their quantization, which is not immediate since the Chern character and the $\hat{A}$-genus are intrinsically rational cohomology classes. To avoid the discussion of these problems \cite{MW}, in the expression \eqref{action} we suppose $C$ to be globally defined, which implies that the field strength $G = dC$ is trivial in the de-Rahm cohomology at any degree.\footnote{Actually the assumption that $C$ is globally defined does not solve the problem, since one should take into account the large gauge transformations $C_{p+1} \rightarrow C_{p+1} + \Phi_{p+1}$ with $\Phi_{p+1}$ integral but not necessarily exact. It turns out that the action \eqref{action} is well-defined under these gauge transformations only under the suitable quantization conditions we have mentioned above. Anyway, for a fixed global $C_{p+1}$ formula \eqref{action} is meaningful, and this is enough for our pourposes here.} For a general discussion see \cite{Freed2}.

\paragraph{} We put for notational convenience:
	\[\textstyle G(WY_{p}) = e^{\frac{d}{2}} \wedge \frac{\,\sqrt{\hat{A}(T(WY_{p}))}\,} {\sqrt{\hat{A}(N(WY_{p}))}} \; .
\]
The action \eqref{action} is equal to:
	\[S = \int_{\PD_{WY_{p}}(\ch(E))} i^{*}C \wedge G(WY_{p}) \; .
\]
Let $\{q_{k} \cdot WY_{k}\}$ be the set of D-branes appearing in the Poincar\'e dual of $\ch(E)$ in $WY_{p}$ (we mean that we choose a representative cycle for each homology class in $\PD_{WY_{p}}(\ch(E))$ and we think of it as a subbrane of $WY_{p}$): the first one is $\PD_{WY_{p}}(\ch_{0}(E)) = q \cdot WY_{p}$, so it gives rise to the action without gauge coupling. The other ones are lower dimensional branes. Let us consider the first one, i.e.\ $WY_{(1)} = \PD_{WY_{p}}(\ch_{1}(E))$. Then the correponding term in the action is $\int_{WY_{(1)}} i^{*}C \wedge G(WY_{p})$, which can be written as $\int_{WY_{(1)}} i^{*}C \wedge G(WY_{(1)}) + \int_{WY_{(1)}} i^{*}C \wedge (G(WY_{p}) - G(WY_{(1)}))$. Since in the second term the sum $G(WY_{p}) - G(WY_{(1)})$ has 0-term equal to $0$, then $\PD_{WY_{(1)}} (G(WY_{p}) - G(WY_{1}))$ is made only by lower-dimensional subbranes. Let $WY_{(1,1)}$ be the first one: we get $\int_{WY_{(1,1)}} i^{*}C$, which is equal to $\int_{WY_{(1,1)}} i^{*}C \wedge G(WY_{(1,1)}) + \int_{WY_{(1)}} i^{*}C \wedge (1 - G(WY_{(1,1)}))$. The second term gives rise only to lower dimensional subbranes. Proceeding inductively until we arrive at D0-branes, whose $G$-term is $1$, we can write:
	\[\int_{WY_{(1)}} i^{*}C \wedge G(WY_{p}) = \sum_{h=0}^{m} \int_{WY_{(1,h)}} i^{*}C \wedge G(WY_{(1,h)})
\]
where, for $h = 0$, $WY_{(1, 0)} = WY_{(1)}$ holds. Proceeding in the same way for every $WY_{(k)}$, we obtain a set of subbranes $\{q_{k,h} \cdot WY_{(k,h)}\}$, which, using only one index, we still denote by $\{q_{k} \cdot WY_{(k)}\}$. Therefore, calling $i_{k}: WY_{(k)} \rightarrow S$ the embedding, we get:
	\[S = \sum_{k} \int_{WY_{(k)}} i_{k}^{*}C \wedge G(WY_{(k)}) \; .
\]
From this expression we see that \emph{the brane $WY_{p}$ with gauge and gravitational couplings is equivalent to the set of sub-branes $WY_{(k)}$ with trivial gauge bundle}. Moreover we now show that the following equality holds:
\begin{equation}\label{SplittingCharge}
	i_{\#} \bigl( \ch(E) \wedge G(WY_{p}) \bigr) = \sum_{k} (i_{k})_{\#} \,G(WY_{(k)})
\end{equation}
i.e.\ the charge densities of the two configurations are the same. In order to prove this, we recall the formulas:
\begin{equation}\label{FormulasGysin}
\begin{split}
	&i_{\#}(\alpha \wedge i^{*}\beta) = i_{\#}(\alpha) \wedge \beta\\
	&\int_{WY_{p}} \alpha = \int_{S} i_{\#}(\alpha)
\end{split}
\end{equation}
for $\alpha \in H^{*}(WY_{p}, \mathbb{Q})$ and $\beta \in H^{*}(S, \mathbb{Q})$. Thus:
	\[\begin{split}
	&\int_{WY_{p}} i^{*}C \wedge \ch(E) \wedge G(WY_{p}) = \int_{S} i_{\#} \bigl[ i^{*}C \wedge \ch(E) \wedge G(WY_{p}) \bigr]\\
	&\qquad\qquad\qquad = \int_{S} C \wedge i_{\#} \bigl( \ch(E) \wedge G(WY_{p}) \bigr)\\
	&\sum_{k} \int_{WY_{p}} i_{k}^{*}C \wedge G(WY_{(k)}) = \sum_{k} \int_{S} (i_{k})_{\#} \bigl[ i_{k}^{*}C \wedge G(WY_{(k)}) \bigr]\\
	&\qquad\qquad\qquad = \int_{S} C \wedge \sum_{k} (i_{k})_{\#} \bigl( G(WY_{(k)}) \bigr) \; .
\end{split}\]
Since the two terms are equal for every form $C$, we get formula \eqref{SplittingCharge}. We thus get:
\begin{quote}
\textbf{Splitting principle:} \emph{a D-brane $WY_{p}$ with gauge bundle is dynamically equivalent to a set of sub-branes $WY_{(k)}$ with trivial gauge bundle, such that the total charge density of the two configurations is the same.}
\end{quote}
The physical interpretation of this conjecture is the phenomenon of tachyon condensation \cite{Sen,Witten,Evslin}: the quantization of strings extending from a brane to an antibrane leads to a tachyonic mode, which represents an instability and generates a process of annihilation of brane and antibrane world-volumes via an RG-flow \cite{APS}, leaving lower dimensional branes. In particular, given a D-brane $WY_{p}$ with gauge bundle $E \rightarrow WY_{p}$, we can write $E = (E - \rk\, E) + \rk\, E$, so that $E - \rk\, E \in \tilde{K}(WY_{p})$, where $\tilde{K}(WY_{p})$ is the reduced K-theory group of $WY_{p}$ \cite{Atiyah}: thus we think of this configuration as a triple made by a D-brane $WZ_{p}$ with gauge bundle $\rk\, E$, a brane $WY_{p}$ with gauge bundle $E$ and an antibrane $\overline{WZ}_{p}$ with gauge bundle $\rk\, E$. By tachyon condensation only $WZ_{p}$ remains (with trivial bundle, i.e.\ only with its own charge), while $WY_{p}$ and $\overline{WZ}_{p}$ annihilate giving rise to lower dimensional branes with trivial bundle, as stated in the splitting principle. Moreover, if we consider a stack of pairs $(WY_{p}, \overline{WY}_{p})$ with gauge bundles $E$ and $F$ respectively, this is equivalent to consider gauge bundles $E \oplus G$ and $F \oplus G$ respectively, since, viewing the factor $G$ as a stack of pairs $(WZ_{p}, \overline{WZ}_{p})$ with the \emph{same} gauge bundle, it happens that by tachyon condensation $WZ_{p}$ and $\overline{WZ}_{p}$ disappear, leaving no other subbranes. This is the physical interpretation of the stable equivalence relation in K-theory. This principle, as we will see, is an inverse of the Sen conjecture, but we will actully use it to show the Sen conjecture in this setting.

\paragraph{Remark:} the splitting principle holds only at rational level, since it involves Chern characters and $\hat{A}$-genus. At integral level, we do not state such a principle.

\subsubsection{K-theory}

Since we are assuming the $H$-flux to vanish, in order not to be Freed-Witten anomalous the D-brane must be spin$^{c}$. Since the whole space-time is spin, in particular also spin$^{c}$, it follows that the normal bundle of the D-brane is spin$^{c}$ too. Therefore we can consider the K-theory Gysin map $i_{!}: K(WY_{p}) \rightarrow K(S)$ \cite{LM}. We recall the differentiable Riemann-Roch theorem \cite{Hirzebruch, OS}:
\begin{equation}\label{RiemannRoch}
	\ch(i_{!}(E)) \wedge \hat{A}(TS) = i_{\#}\bigl(\ch(E) \wedge e^{\frac{d}{2}} \wedge \hat{A}(T(WY_{p}))\bigr) \; .
\end{equation}
Using \eqref{RiemannRoch} and \eqref{FormulasGysin} we obtain:
\[\begin{split}
	\int_{WY_{p}} i^{*}C \wedge &\,\ch(E) \wedge e^{\frac{d}{2}} \wedge \textstyle \frac{\,\sqrt{\hat{A}(T(WY_{p}))}\,} {\sqrt{\hat{A}(N(WY_{p}))}} \displaystyle = \int_{S} C \wedge \ch\bigl(i_{!}(E)\bigr) \wedge \sqrt{\hat{A}(TS)} \; .
\end{split}\]
Thus we get:
	\[S = \int_{S} C \wedge \ch\bigl(i_{!}(E)\bigr) \wedge \sqrt{\hat{A}(TS)}
\]
hence:
\begin{equation}\label{QWYGysin}
	Q_{WY_{p}} = \ch(i_{!}E) \wedge \sqrt{\hat{A}(TS)} \; .
\end{equation}
In this way, \eqref{QWYGysin} is another expression for $Q_{WY_{p}}$ with respect to \eqref{QWY}, but with an important difference: the $\hat{A}$-factor does not depend on $WY_{p}$, hence all $Q_{WY_{p}}$ is a function only of $E$. Thus, we can consider $i_{!}E$ as the K-theory analogue of the charge density, considered as an \emph{integral} K-theory class. The use of Chern characters, instead, obliges to consider rational classes, which cannot contain information about the torsion part.
 
\subsubsection{Sen conjecture}

Let us consider the two expressions found for the rational charge density:
	\[\begin{split}
	&Q^{(1)}_{WY_{p}} = i_{\#} \bigl( \ch(E) \wedge G(WY_{p}) \bigr) \\
	&Q^{(2)}_{WY_{p}} = \ch(i_{!}E) \wedge \sqrt{\hat{A}(TS)} \; .
\end{split}\]
$Q^{(2)}_{WY_{p}}$ is exactly the charge density of a stack of D$9$-branes and anti-branes (whose world-volume coincides with $S$), whose gauge bundle is the K-theory class $i_{!}E$. Hence, expressing the charge in the form $Q^{(2)}_{WY_{p}}$ for each D-brane in our background is equivalent to think that there exists only one stack of couples brane-antibrane of dimension $9$ encoding all the dynamics. Hence we formulate the conjecture \cite{Sen,Witten}:
\begin{quote}
\textbf{Sen conjecture:} \emph{every configuration of branes and anti-branes with any gauge bundle is dynamically equivalent to a configuration with only a stack of coincident pairs brane-antibrane of dimension $9$ with an appropriate K-theory class on it.}
\end{quote}
In order to see that the dynamics is actually equivalent, we use the splitting principle stated above: since $Q^{(1)}_{WY_{p}} = Q^{(2)}_{WY_{p}}$, the brane $WY_{p}$ with the charge $Q^{(1)}_{WY_{p}}$ and the D$9$-brane with charge $Q^{(2)}_{WY_{p}}$ split into the same set of subbranes (with trivial gauge bundle). We remark that in order to state the Sen conjecture is necessary that the $H$-flux vanishes. Indeed, the space-time is spin$^{c}$ (it is spin since space-time spinors exist, therefore also spin$^{c}$), hence Freed-Witten anomaly cancellation for D$9$-branes requires that $H = 0$. Actually, an appropriate stack of D$9$-branes can be consistent for $H$ a torsion class \cite{Kapustin}, but we do not consider this case in the present paper.

\paragraph{}In order to formulate both the splitting principle and the Sen conjecture, we have only considered the action, hence only \emph{rational} classes given by Chern characters and $\hat{A}$-genus. Thus, we can classify the charge density in the two following ways:
\begin{itemize}
	\item as a rational cohomology class $i_{\#} (\ch(E) \wedge G(WY_{p})) \in H^{\ev}(S, \mathbb{Q})$;
	\item as a rational K-theory class $i_{!}E \in K_{\mathbb{Q}}(S) := K(S) \otimes_{\mathbb{Z}} \mathbb{Q}$.
\end{itemize}
These two classification schemes are completely equivalent due to the fact that the map:
	\[\ch(\,\cdot\,) \wedge \sqrt{\hat{A}(TS)}:\; K_{\mathbb{Q}}(S) \longrightarrow H^{\ev}(S, \mathbb{Q})
\]
is an isomorphism. This equivalence is lost at the integral level, since the torsion parts of $K(S)$ and $H^{\ev}(S, \mathbb{Z})$ are in general different. Moreover, since at the integral level the splitting principle does not apply, we cannot prove that the Sen conjecture holds: the classification via Gysin map and cohomology are different, and the use of the Gysin map is just \emph{suggested} by the equivalence at rational level, i.e.\ by the equivalence of the dynamics.

Moreover, for the integral case, we have also seen the classification via the Atiyah-Hirzebruch spectral sequence (AHSS). In the rational case, we can build the corresponding sequence AHSS$_{\mathbb{Q}}$ \cite{AH}, ending at the groups $Q_{\infty}^{\ev, \,0}(S)$, but it stabilizes at the second step, i.e.\ at the level of cohomology. Hence, the class $\{i_{\#}(\ch(E) \wedge G(WY_{p}))\} \in Q^{\ev,\,0}_{\infty}(S)$ is completely equivalent to the cohomology class $i_{\#}(\ch(E) \wedge G(WY_{p})) \in H^{\ev}(S, \mathbb{Q})$.

\subsection{Linking the classifications}

To summarize, we are trying to classify the charges of D-branes in a compact euclidean space-time $S$. In order to achieve this, we can use cohomology or K-theory, with integer or rational coefficients, obtaining the possibilities showed in table \ref{fig:Classifications}.
\begin{table*}[h]
	\centering
		\begin{tabular}{|l|l|l|}
			\hline & & \\ & Integer & Rational \\ & & \\ \hline
			& & \\ Cohomology & $\PD_{S}[q \cdot WY_{p}] \in H^{9-p}(S, \mathbb{Z})$ & $i_{\#}\bigl(\ch(E) \wedge G(WY_{p})\bigr) \in H^{\ev}(S, \mathbb{Q})$ \\ & & \\ \hline
			& & \\ K-theory (Gysin map) & $i_{!}(E) \in K(S)$ & $i_{!}(E) \in K_{\mathbb{Q}}(S)$ \\ & & \\ \hline
			& & \\ K-theory (AHSS) & $\{\PD_{S}[q\cdot WY_{p}]\} \in E^{9-p,\,0}_{\infty}(S)$ & $\bigl\{i_{\#}(\ch(E) \wedge G(WY_{p}))\bigr\} \in Q^{\ev,\,0}_{\infty}(S)$ \\ & & \\ \hline
		\end{tabular}
\caption{Classifications}\label{fig:Classifications}
\end{table*}

In the rational case, as we have seen, there is a complete equivalence of the three approaches, since the three groups we consider, i.e.\ $\bigoplus_{2k}H^{2k}(S, \mathbb{Q})$, $K_{\mathbb{Q}}(S)$ and $\bigoplus_{2k}Q^{2k,\,0}_{\infty}(S)$ are canonically isomorphic. Instead, in the integral case there are no such isomorphisms (in general the three groups are all different), and there is a strong asymmetry due to the fact that in the homological and AHSS classifications \emph{the gauge bundle and the gravitational coupling are not considered at all}, while they are of course taken into account in the Gysin map approach. Up to now we discussed the case of even-codimensional branes: that is because the Gysin map requires an even-dimensional normal bundle in order to take value in $K(S)$. We will discuss also the odd-dimensional case, considering the group $K^{1}(S)$, and the picture will be similar.

Since the integral approaches are not equivalent, we have to investigate the relations among them: it is clear how to link the cohomology class and the AHSS class, since the second step of AHSS is exactly the cohomology. Our aim is to find an explicit link between the Gysin map approach and the one based on AHSS.

\section{Useful notions of K-theory}

We briefly recall the main K-theoretical constructions which will be used in the following. In this section we use the following notations: $X$ and $Y$ are topological spaces, $K(X)$ is the K-theory group of $X$, $\tilde{K}(X)$ is the reduced K-theory group of $X$, $K^{n}(X)$ is the K-theory group of degree $n$ of $X$ and $\tilde{K}^{n}(X)$ is the reduced K-theory group of degree $n$ of $X$ \cite{Atiyah, LM}. If $f: X \rightarrow Y$ is a continuous map and $E$ is a vector bundle on $Y$, we denote by $f^{*}E$ the pull-back of $E$ on $X$; if $\alpha = [E] - [F]$ is a K-theory class on $Y$, we denote by $f^{*}\alpha$ its pull-back $f^{*}\alpha = [f^{*}E] - [f^{*}F]$. Moreover:
\begin{itemize}
	\item fixing two marked points $x_{0} \in X$ and $y_{0} \in Y$, we put $X \vee Y := (\{x_{0}\} \times Y) \cup (X \times \{y_{0}\})$ and $X \wedge Y := (X \times Y)/(X \vee Y)$;
	\item we denote by $X^{+}$ the one-point compactification of $X$ \cite{Dugundji}. We call $\{\infty\}$ the point added in such a compactification.
\end{itemize}

\subsection{Products in K-theory}

For $X$ a topological space, $K(X)$ has a natural ring structure given by the tensor product: $[E] \otimes [F] := [E \otimes F]$. Such a product restricts to $\tilde{K}(X)$. In general, we can define a product:
\begin{equation}\label{SquareProduct}
	K(X) \otimes K(Y) \overset{\boxtimes}\longrightarrow K(X \times Y)
\end{equation}
where, if $\pi_{1}: X \times Y \rightarrow X$ and $\pi_{2}: X \times Y \rightarrow Y$ are the projections, $E \boxtimes F = \pi_{1}^{*}E \otimes \pi_{2}^{*}F$. The fiber of $E \boxtimes F$ at $(x,y)$ is $E_{x} \otimes E_{y}$.\footnote{If $X = Y$ and $\Delta: X \rightarrow X \times X$ is the diagonal embedding, then $E \otimes F = \Delta^{*}(E \boxtimes F)$.} We now prove that, fixing a marked point for $X$ and $Y$, the product \eqref{SquareProduct} restricts to (see \cite{Piazza}):
\begin{equation}\label{ProductWedge}
	\tilde{K}(X) \otimes \tilde{K}(Y) \overset{\boxtimes}\longrightarrow \tilde{K}(X \wedge Y) \; .
\end{equation}
For this, we first state that:\footnote{\eqref{TimesSplitting} is actually true for $\tilde{K}^{n}(X \times Y)$ for any $n$, with the same proof.}
\begin{equation}\label{TimesSplitting}
	\tilde{K}(X \times Y) \simeq \tilde{K}(X \wedge Y) \oplus \tilde{K}(Y) \oplus \tilde{K}(X) \; .
\end{equation}
In fact:
\begin{itemize}
	\item since $X$ is a retract of $X \times Y$ via the projection, we have that $\tilde{K}(X \times Y) = K(X \times Y, X) \oplus \tilde{K}(X) = \tilde{K}(X \times Y / X) \oplus \tilde{K}(X)$ (see \cite{Atiyah});
	\item since $Y$ is a retract of $X \times Y / X$ via the projection, we also have $\tilde{K}(X \times Y / X) = K(X \times Y / X, \, Y) \oplus \tilde{K}(Y) = \tilde{K}(X \wedge Y) \oplus \tilde{K}(Y)$.
\end{itemize}
Combining we obtain \eqref{TimesSplitting}. We describe the explicit isomorphism. We call $i_{1}: X \rightarrow X \times Y$ and $i_{2}: Y \rightarrow X \times Y$ the immersions defined by $i_{1}(x) = (x, y_{0})$ and $i_{2}(y) = (x_{0}, y)$, and, for $\alpha \in K(X \times Y)$, we put $\alpha\vert_{X} := (i_{1})^{*}\alpha$ and $\alpha\vert_{Y} := (i_{2})^{*}\alpha$. Then, for $\alpha = [E] - [F] \in \tilde{K}(X \times Y)$, the explicit isomorphism in \eqref{TimesSplitting} is:
	\[\alpha \longrightarrow \bigl(\alpha - \pi_{1}^{*}\, (\alpha\vert_{X}) - \pi_{2}^{*}\, (\alpha\vert_{Y}) \bigr) \oplus \, \alpha\vert_{Y} \oplus\, \alpha\vert_{X} \; .
\]
Let $\alpha \in \tilde{K}(X)$ and $\beta \in \tilde{K}(Y)$: then $(\alpha \boxtimes \beta) \vert_{X} = 0$ and $(\alpha \boxtimes \beta) \vert_{Y} = 0$. In fact,  one has:
	\[(\alpha \boxtimes \beta) \vert_{X} = \alpha \otimes (\pi_{2}^{*} \,\beta)\vert_{X} = \alpha \otimes i_{1}^{*}\pi_{2}^{*} \,\beta = \alpha \otimes (\pi_{2}i_{1})^{*} \,\beta \; .
\]
But $\pi_{2}i_{1}: X \rightarrow Y$ is the constant map with value $y_{0}$, and the pull-back of a bundle by a constant map is trivial: since $\beta$ is a reduced K-theory class, it follows that $(\pi_{2}i_{1})^{*} \,\beta = 0$. Similarly for $Y$. Hence, by \eqref{TimesSplitting}, we obtain that $\alpha \boxtimes \beta \in \tilde{K}(X \wedge Y)$.

\subsubsection{Non-compact case}

For a generic (also non-compact) space $X$, we use K-theory with compact support, i.e.\ we define $K(X) := \tilde{K}(X^{+})$ (for compact spaces this definition coincides with the usual one up to canonical isomorphism). One can easily prove that $X^{+} \wedge Y^{+} = (X \times Y)^{+}$, considering as marked points on $X$ and $Y$ the points at infinity. Hence, the product \eqref{ProductWedge} becomes exactly:
\begin{equation}\label{ProductNonCompact}
	K(X) \otimes K(Y) \overset{\boxtimes}\longrightarrow K(X \times Y)
\end{equation}
also for the non-compact case.

\subsection{Thom isomorphism}

Let $X$ be a \emph{compact} topological space and $\pi: E \rightarrow X$ a vector bundle (real or complex): we show that $K(E)$ has a natural structure of $K(X)$-module. It seems natural to use the pull-back $\pi^{*}: K(X) \rightarrow K(E)$, but this is not possible: in fact, the group $K(E)$ is defined as the reduced K-theory group of $E^{+}$, and in general there are no possibilities to extend continuously the projection $\pi$ to $E^{+}$. Hence we use the product \eqref{ProductNonCompact}: considering the embedding $i: E \rightarrow X \times E$ defined by $i(e) = (\pi(e), e)$,\footnote{For such an embedding it is not necessary to have a marked point on $X$.} which trivially extends to $i: E^{+} \rightarrow (X \times E)^{+}$ requiring that $i(\infty) = \infty$, we can define a product:
\begin{equation}\label{ProductModule}
\begin{split}
	K(&X) \otimes K(E) \longrightarrow K(E)\\
	&\alpha \otimes \beta \longrightarrow i^{*}(\alpha \boxtimes \beta).
\end{split}
\end{equation}
This product defines a structure of $K(X)$-module on $K(E)$.

\begin{Lemma}\label{Unitarity} $K(E)$ is \emph{unitary} as a $K(X)$-module.
\end{Lemma}
\textbf{Proof:} Let us consider the following maps:
	\[\begin{split}
	&\pi_{1}: X^{+} \times E^{+} \longrightarrow X^{+}\\
	&\pi_{2}: X^{+} \times E^{+} \longrightarrow E^{+}\\
	&i: E^{+} \longrightarrow (X \times E)^{+}\\
	&\tilde{\pi}: X^{+} \times E^{+} \longrightarrow X^{+} \wedge E^{+} = (X \times E)^{+}\\
	&\tilde{\pi}_{2}: (X \times E)^{+} \longrightarrow E^{+}
\end{split}\]
where $i(e) = (\pi(e), e)$ and the others are defined in the obvious way. Since the map:
	\[r: X^{+} \times E^{+} \longrightarrow \bigl(X^{+} \times \{\infty\}\bigr) \cup \bigl(\{\infty\} \times E^{+}\bigr)
\]
given by $r(x, e) = (x, \infty)$ and $r(\infty, e) = (\infty, e)$ \footnote{The map $r$ is continuous because $X$ is compact, so that its $\infty$-point is disjoint from it.} is a retraction, $\tilde{\pi}^{*}: \tilde{K}((X \times E)^{+}) \rightarrow \tilde{K}(X^{+} \times E^{+})$ is injective \cite{Atiyah}. Then, by the definition of the module structure, for $\alpha \in K(X) = \tilde{K}(X^{+})$ and $\beta \in K(E) = \tilde{K}(E^{+})$ we reformulate \eqref{ProductModule} as:\footnote{With respect to \eqref{ProductModule} we think $\alpha \boxtimes \beta \in \tilde{K}(X^{+} \times E^{+})$ and we write explicitly $(\tilde{\pi}^{*})^{-1}$.}
	\[\alpha \cdot \beta = i^{*} (\tilde{\pi}^{*})^{-1}(\alpha \boxtimes \beta) = i^{*} (\tilde{\pi}^{*})^{-1} (\pi_{1}^{*}\alpha \otimes \pi_{2}^{*}\beta) \; .
\]
For $\alpha = 1$ one has $\alpha\vert_{X} = X \times \mathbb{C}$ and $\alpha\vert_{\{\infty\}} = 0$. Hence:
	\[\begin{split}
	&(1 \boxtimes \beta)\,\big\vert_{X \times E^{+}} = \pi_{2}^{*}\beta\,\big\vert_{X \times E^{+}}\\
	&(1 \boxtimes \beta)\,\big\vert_{\{\infty\} \times E^{+}} = 0.
\end{split}\]
But:
\begin{itemize}
	\item since $\pi_{2} \,\big\vert_{X \times E^{+}} = (\tilde{\pi}_{2} \circ \tilde{\pi}) \,\big\vert_{X \times E^{+}}$, one has $\pi_{2}^{*}\beta\,\big\vert_{X \times E^{+}} = \tilde{\pi}^{*}\tilde{\pi}_{2}^{*} \beta\,\big\vert_{X \times E^{+}}$;
	\item since $\tilde{\pi}_{2} \circ\, \tilde{\pi}\, (\{\infty\} \times E^{+}) = \{\infty\}$ and $\beta \in \tilde{K}(E^{+})$, one has $(\tilde{\pi}^{*}\tilde{\pi}_{2}^{*} \beta) \, \big\vert_{\{\infty\} \times E^{+}} = 0$.
\end{itemize}
Hence $1 \boxtimes \beta = \tilde{\pi}^{*}\tilde{\pi}_{2}^{*} \beta$, so that:
	\[1 \cdot \beta = i^{*} (\tilde{\pi}^{*})^{-1}\tilde{\pi}^{*}\tilde{\pi}_{2}^{*} \beta = i^{*} \tilde{\pi}_{2}^{*} \beta = (\tilde{\pi}_{2} \circ i)^{*}\beta = \id^{*}\beta = \beta.
\]
$\square$

\paragraph{}Let us consider a vector space $\mathbb{R}^{2n}$ as a vector bundle on a point $\{x\}$. Then we have:
\begin{itemize}
	\item $K(\{x\}) = \mathbb{Z}$;
	\item $K(\mathbb{R}^{2n}) = \tilde{K}((\mathbb{R}^{2n})^{+}) = \tilde{K}(S^{2n}) = \mathbb{Z}$.
\end{itemize}
Hence $K(\{x\}) \simeq K(\mathbb{R}^{2n})$. The idea of the Thom isomorphism is to extend this isomorphism to a generic bundle $E \rightarrow X$ with fiber $\mathbb{R}^{2n}$. To achieve this, we try to write such an isomorphism in a way that extends to a generic bundle. Actually, this generalization works for $E$ a spin$^{c}$-bundle of even dimension.

Let us consider the spin group $\Spin(2n)$ \cite{LM}. The spin representation acts on $\mathbb{C}^{2^{n}}$, and it splits in the two irreducible representations of positive and negative chirality, acting on the subspaces $S^{+}$ and $S^{-}$ of $\mathbb{C}^{2^{n}}$ of dimension $2^{n-1}$. Also the group $\Spin^{c}(2n)$, defined as $\Spin(2n) \otimes_{\mathbb{Z}_{2}} U(1)$, acts on $\mathbb{C}^{2^{n}}$ via the standard spin$^{c}$ representation, and the same splitting in chirality holds: we call the two corresponding subspaces $S^{+}_{\mathbb{C}}$ and $S^{-}_{\mathbb{C}}$ when we think of them as $\Spin^{c}(2n)$-modules instead of $\Spin(2n)$-modules. For $\CCl(2n)$ the complex Clifford algebra of dimension $2n$, $\mathbb{C}^{2^{n}}$ is also a $\CCl(2n)$-module, and, for $v \in \mathbb{R}^{2n} \subset \CCl(2n)$, we have $v \cdot S_{\mathbb{C}}^{+} = S_{\mathbb{C}}^{-}$. We thus consider the following complex:
	\[0 \longrightarrow \mathbb{R}^{2n} \times S_{\mathbb{C}}^{+} \overset{c}\longrightarrow \mathbb{R}^{2n} \times S_{\mathbb{C}}^{-} \longrightarrow 0
\]
where $c$ is the Clifford multiplication by the first component: $c(v, z) = (v, v \cdot z)$. Such a sequence of trivial bundles on $\mathbb{R}^{2n}$ is exact when restricted to $\mathbb{R}^{2n} \setminus \{0\}$, hence the alternated sum:
	\[\lambda_{\mathbb{R}^{2n}} = \bigl[ \mathbb{R}^{2n} \times S_{\mathbb{C}}^{-} \bigr] - \bigl[ \mathbb{R}^{2n} \times S_{\mathbb{C}}^{+} \bigr]
\]
naturally gives a class in $K(\mathbb{R}^{2n}, \mathbb{R}^{2n} \setminus \{0\})$  \cite{Atiyah}. The sequence is exact in particular in $\mathbb{R}^{2n} \setminus B^{2n}$, where $B^{2n}$ is the open ball of radius $1$ in $\mathbb{R}^{2n}$, hence it defines a class:
	\[\lambda_{\mathbb{R}^{2n}} \in K(\mathbb{R}^{2n}, \mathbb{R}^{2n} \setminus B^{2n}) = \tilde{K}( \overline{B^{2n}} / S^{2n-1} ) = \tilde{K}(S^{2n}) \; .
\]
One can prove that, for $\eta$ the dual of the tautological line bundle on $\mathbb{CP}^{1}$, whose sheaf of sections is usually denoted as $\mathcal{O}_{\mathbb{CP}^{1}}(1)$, if we identify $S^{2}$ with $\mathbb{CP}^{1}$ topologically, we have that:
\begin{equation}\label{LambdaR2n}
	\lambda_{\mathbb{R}^{2n}} = (-1)^{n} \cdot (\eta - 1)^{\boxtimes n}
\end{equation}
i.e.\ $\lambda_{\mathbb{R}^{2n}}$ is a generator of $\tilde{K}(S^{2n}) \simeq \mathbb{Z}$ \cite{Atiyah}.

We now show the generalization to a spin$^{c}$-bundle $\pi: E \rightarrow X$ of dimension $2n$. Let $S^{\pm}_{\mathbb{C}}(E)$ be the bundles of complex chiral spinors associated to $E$: to define them, we consider a spin$^{c}$-lift of the orthogonal frame bundle $\SO(E)$, which we call $\Spin^{c}(E)$, and we define $S_{\mathbb{C}}(E)$ as the vector bundle with fiber $\mathbb{C}^{2^{n}}$ associated to the spin$^{c}$ representation, the latter being induced by the action of the complex Clifford algebra via the inclusion $\Spin^{c}(2n) \subset \CCl(2n) \hookrightarrow \mathbb{C}^{2^{n}}$. This bundle splits into $S_{\mathbb{C}}(E) = S^{+}_{\mathbb{C}}(E) \oplus S^{-}_{\mathbb{C}}(E)$; moreover, $S_{\mathbb{C}}(E)$ is naturally a $\CCl(E)$-module. We can lift $S^{\pm}_{\mathbb{C}}(E)$ to $E$ by $\pi^{*}$. Then we consider the complex:
	\[0 \longrightarrow \pi^{*}S^{+}_{\mathbb{C}}(E) \overset{c}\longrightarrow \pi^{*}S^{-}_{\mathbb{C}}(E) \longrightarrow 0
\]
where $c$ is the Clifford multiplication given by the structure of $\CCl(E)$-module: for $e \in E$ and $s_{e} \in (\pi^{*}S^{+}_{\mathbb{C}}(E))_{e}$, we define $c(s_{e}) = e \cdot s_{e}$. Such a sequence is exact when restricted to $E \setminus B(E)$, where, for any fixed metric on $E$, $B(E)$ is the union of the open balls of radius $1$ on each fiber. Hence we can define the \emph{Thom class}:
\begin{equation}\label{ThomClass}
	\lambda_{E} = [\pi^{*}S^{-}_{\mathbb{C}}(E)] - [\pi^{*}S^{+}_{\mathbb{C}}(E)]
\end{equation}
as a class in $K(\,E ,\, E \setminus B(E)\,) = \tilde{K}( \, \overline{B(E)} \,/\, S(E) \, ) = \tilde{K}(E^{+}) = K(E)$. The following fundamental theorem holds (\cite{LM,Karoubi} and, only for the complex case, \cite{Atiyah, Piazza}):

\begin{Theorem}[Thom isomorphism] Let $X$ be a compact topological space and $\pi: E \rightarrow X$ an \emph{even} dimensional spin$^{c}$-bundle. For
	\[\lambda_{E} = [\pi^{*}S^{-}_{\mathbb{C}}(E)] - [\pi^{*}S^{+}_{\mathbb{C}}(E)] \in K(E)
\]
the map, defined using the module structure \eqref{ProductModule}:
	\[\begin{split}
	T: \; &K(X) \longrightarrow K(E)\\
	& \alpha \rightarrow \alpha \cdot \lambda_{E}
\end{split}\]
is a group isomorphism.
\end{Theorem}

\paragraph{}We can now see that the construction for a generic $2n$-dimensional spin$^{c}$-bundle $E \rightarrow X$ is a generalization of the construction for $\mathbb{R}^{2n}$. In fact, for $x \in X$:
\begin{itemize}
	\item $\bigl( \pi^{*}S^{\pm}_{\mathbb{C}}(E) \bigr) \big\vert_{E_{x}} = E_{x} \times \bigl( S^{\pm}_{\mathbb{C}}(E) \bigr)_{x} \simeq \mathbb{R}^{2n} \times S^{\pm}_{\mathbb{C}}(\mathbb{R}^{2n})$;
	\item the Clifford multiplication restricts on each fiber $E_{x}$ to the Clifford multiplication in $\mathbb{R}^{2n} \times S_{\mathbb{C}}(\mathbb{R}^{2n})$.
\end{itemize}
Hence:
\begin{equation}\label{RestrictionThom}
\lambda_{E} \big\vert_{E_{x}} \simeq \lambda_{\mathbb{R}^{2n}} \; .
\end{equation}
In particular, we see that, for $i: E_{x}^{+} \rightarrow E^{+}$, the restriction $i^{*}: K(E) \rightarrow K(E_{x}) \simeq \mathbb{Z}$ is surjective.

\subsection{Gysin map}

Let $X$ be a compact smooth $n$-manifold and $Y \subset X$ a compact embedded $p$-submanifold such that $n - p$ is even and the normal bundle $\mathcal{N}(Y) = (TX\,\vert_{Y}) / \,TY$ is spin$^{c}$. Then, since $Y$ is compact, there exists a tubular neighborhood $U$ of $Y$ in $X$, i.e.\ there exists an homeomorphism $\varphi_{U}: U \rightarrow \mathcal{N}(Y)$.

If $i: Y \rightarrow X$ is the embedding, from this data we can naturally define a group homomorphism, called \emph{Gysin map}:
	\[i_{!}: K(Y) \longrightarrow \tilde{K}(X) \; .
\]
In fact:
\begin{itemize}
	\item we first apply the Thom isomorphism $T: K(Y) \longrightarrow K(\mathcal{N}(Y)) = \tilde{K}(\mathcal{N}(Y)^{+})$;
	\item then we naturally extend $\varphi_{U}$ to $\varphi_{U}^{+}: U^{+} \longrightarrow \mathcal{N}(Y)^{+}$ and apply $(\varphi_{U}^{+})^{*}: K(\mathcal{N}(Y)) \longrightarrow K(U)$;
	\item there is a natural map $\psi: X \rightarrow U^{+}$ defined by:
	\[\psi(x) = \left\{\begin{array}{ll}
	x & \text{if } x \in U \\
	\infty & \text{if } x \in X \setminus U
	\end{array}\right.
\]
hence we apply $\psi^{*}: K(U) \rightarrow \tilde{K}(X)$.
\end{itemize}
Summarizing:
\begin{equation}\label{GysinMap}
	i_{!}\,(\alpha) = \psi^{*} \circ (\varphi_{U}^{+})^{*} \circ T \, (\alpha) \; .
\end{equation}

\paragraph{Remark:} One could try to use the immersion $i: U^{+} \rightarrow X^{+}$ and the retraction $r: X^{+} \rightarrow U^{+}$ to have a splitting $K(X) = K(U) \oplus K(X, U) = K(Y) \oplus K(X,U)$. This is false, since the immersion $i: U^{+} \rightarrow X^{+}$ is not continuous: \emph{since $X$ is compact}, $\{\infty\} \subset X^{+}$ is open, but $i^{-1}(\{\infty\}) = \{\infty\}$, and $\{\infty\}$ is not open in $U^{+}$ since $U$ is not compact.


\section{The Atiyah-Hirzebruch spectral sequence}

We recall how to construct the Atiyah-Hirzebruch spectral sequence, and we introduce the tools we need in order to link it with the Gysin map.

\subsection{Spectral sequence for a cohomology theory}

We deal with spectral sequences in the axiomatic version described in \cite{CE}, chap.\ XV, par.\ 7, with the additional hypotesis of working with \emph{finite} sequences of groups. We also take into account the presence of the grading in cohomology. In particular, we suppose the following assignements are given for $p, p', p'' \in \mathbb{Z} \cup \{-\infty, +\infty\}$:
\begin{itemize}
	\item for $-\infty \leq p \leq p' \leq \infty$, abelian groups $H^{n}(p,p')$ for $n \in \mathbb{Z}$, such that $H^{n}(p, p') = H^{n}(0, p')$ for $p \leq 0$ and there exists $l \in \mathbb{N}$ such that $H^{n}(p, p') = H^{n}(p, +\infty)$ for $p' > l$ ($l$ does not depend on $n$ in our setting);
	\item for $p \leq p' \leq p''$, $a, b \geq 0$, $p+a \leq p'+b$, two maps:\footnote{The map $\delta$ is called in the same way in \cite{CE}. Instead, we introduce the name $\psi$ since the analogous map in \cite{CE} has no name.}
\begin{equation}\label{PsiDelta}
\begin{split}
	&\psi^{n}: H^{n}(p+a,p'+b) \rightarrow H^{n}(p,p')\\
	&\delta^{n}: H^{n}(p,p') \rightarrow H^{n+1}(p',p'')
\end{split}
\end{equation}
\end{itemize}
satisfying axioms (SP.1)-(SP.5) of \cite{CE}, p.\ 334. When the indices are not clear from the context, we also use the notations $(\psi^{n})^{p+a,p'+b}_{p,p'}$ and $(\delta^{n})^{p,p',p''}$ for the maps \eqref{PsiDelta}. We can describe the groups and the coboundaries of the spectral sequence in the following way:
\begin{equation}\label{Epr}
\begin{array}{ll}
	E^{p,\,q}_{r} = \IIm\bigl( H^{p+q}(p, p+r) \overset{\psi^{p+q}} \longrightarrow H^{p+q}(p-r+1, p+1) \bigr) & \textnormal{(\cite{CE}, formula (8) p.\ 318)} \\ \\
	d^{p,\,q}_{r} \,=\, (\delta^{p+q})^{p-r+1,p+1,p+r+1} \,\big\vert_{\IIm((\psi^{p+q})^{p,p+r}_{p-r+1,p+1})}\,:\\ \\
	\phantom{XXXXXXXXXXXXXXXXXX} E^{p,q}_{r} \longrightarrow E^{p+r,q-r+1}_{r} & \textnormal{(\cite{CE}, line 3 p.\ 319)} \\ \\
	F^{p,\,q}H = \IIm \bigl( H^{p+q}(p, +\infty) \overset{\psi^{p+q}}\longrightarrow H^{p+q}(0, +\infty) \bigr) & \textnormal{(\cite{CE}, line -10 p.\ 319)} \; .
\end{array}
\end{equation}
Then:
\begin{itemize}
	\item the groups $F^{p,\,q}H$ are a filtration of $H^{p+q}(0, +\infty)$;
	\item $\bigoplus_{p,q} E^{p,\,q}_{r+1} \simeq H\bigl( \bigoplus_{p,q} E^{p,\,q}_{r}, \bigoplus_{p,q} d^{p,\,q}_{r} \bigr)$ canonically, i.e.\ $E^{p,\,q}_{r+1} \simeq \Ker \, d^{p,\,q}_{r} / \IIm \, d^{p-r, \,q+r-1}_{r}$;
	\item the sequence $\{E^{p,\,q}_{r}\}_{r \in \mathbb{N}}$ stabilizes to $F^{p,\,q}H / F^{p+1,\,q-1}H$.
\end{itemize}
In particular, considering the following commutative diagram\footnote{The maps $\psi_{1}, \psi_{2}, \delta_{1}, \delta_{2}$ of the diagram are maps of the family \eqref{PsiDelta}; here and in the following we use this notation in order not to write too many indices.} (\cite{CE}, end of p.\ 318):
\begin{equation}\label{BoundaryDiagram}
\xymatrix{
H^{p+q}(p, p+r) \ar[r]^{\psi_{1} \quad\;\;} \ar[d]_{\delta_{1}} & H^{p+q}(p-r+1, p+1) \ar[d]^{\delta_{2}}\\
H^{p+q+1}(p+r, p+2r) \ar[r]^{\psi_{2} \;\;} & H^{p+q+1}(p+1,p+r+1)
}
\end{equation}
the following identities hold:
\begin{itemize}
	\item $\IIm(\psi_{1}) = E^{p,\,q}_{r}$ and $\IIm(\psi_{2}) = E^{p+r,\,q-r+1}_{r}$;
	\item $d^{p,\,q}_{r} = \delta_{2} \,\big\vert_{\IIm(\psi_{1})}\,:\, E^{p,\,q}_{r} \rightarrow E^{p+r,\,q-r+1}_{r}$.
\end{itemize}
The limit of the sequence $\bigoplus_{p} F^{p,\,q}H / F^{p+1,\,q-1}H$ can also be defined as (\cite{CE}, eq.\ (3) p.\ 316):
\begin{equation}\label{Limit}
	E^{p,\,q}_{0}H := E^{p,\,q}_{\infty} = \IIm \bigl( H^{p+q}(p, +\infty) \overset{\psi^{p+q}}\longrightarrow H^{p+q}(0, p+1) \bigr)
\end{equation}
i.e.\ $E^{p,\,q}_{0}H \simeq F^{p,\,q}H / F^{p+1,\,q-1}H$ canonically.

\paragraph{}Given a topological space $X$ with a finite filtration:
	\[\emptyset = X^{-1} \subset X^{0} \subset \cdots \subset X^{m} = X
\]
we can consider a cohomology theory $h^{\bullet}$ \cite{Hatcher} and define (for $p \leq p' \leq p''; a, b \geq 0; p+a \leq p'+b$):
\begin{itemize}
	\item $H^{n}(p,p') = h^{n}(X^{p'-1}, X^{p-1})$;
	\item $\psi^{n}: H^{n}(p+a, p'+b) \rightarrow H^{n}(p,p')$ is induced (thanks to the axioms of cohomology) by the map of couples $i: (X^{p'-1}, X^{p-1}) \rightarrow (X^{p'+b-1}, X^{p+a-1})$;
	\item $\delta^{n}: H^{n}(p,p') \rightarrow H^{n}(p',p'')$ is the composition of the map $\pi^{*}: h^{n}(X^{p'-1}, X^{p-1}) \rightarrow h^{n}(X^{p'-1})$ induced by the map of couples $\pi: (X^{p'-1}, \emptyset) \rightarrow (X^{p'-1}, X^{p-1})$, and the Bockstein map $\beta^{n}: h^{n}(X^{p'-1}) \rightarrow h^{n+1}(X^{p''-1},X^{p'-1})$.
\end{itemize}

\paragraph{Remark:} the shift by $-1$ in the definition of $H^{n}(p,p')$ is necessary in order to have the equality $H^{n}(0, +\infty) = h^{n}(X)$. It would not be necessary if we decleared $X^{0} = \emptyset$, but this is not coherent with the case of simplicial complexes, since, in that case, $X^{0}$ denotes the $0$-skeleton.

\paragraph{}Since K-theory is a cohomology theory, it is natural to consider the spectral sequence associated to it for a given filtration $\emptyset = X^{-1} \subset X^{0} \subset \cdots \subset X^{m} = X$: such a sequence is called Atiyah-Hirzebruch spectral sequence (AHSS). In particular, groups and maps are defined in the following way (for $p \leq p' \leq p''; a, b \geq 0; p+a \leq p'+b$):
\begin{itemize}
	\item $H^{n}(p,p') = K^{n}(X^{p'-1}, X^{p-1})$;
	\item $\psi^{n}: K^{n}(X^{p'+b-1}, X^{p+a-1}) \rightarrow K^{n}(X^{p'-1}, X^{p-1})$ is the pull-back via the map $i: X^{p'-1}/ X^{p-1}$ $\rightarrow X^{p'+b-1} / X^{p+a-1}$ (we recall that, for spaces having the homotopy type of a finite simplicial complex, $K^{n}(X, Y) = \tilde{K}^{n}(X/Y)$ by definition \cite{Atiyah});
	\item $\delta^{n}: K^{n}(X^{p'-1}, X^{p-1}) \longrightarrow K^{n+1}(X^{p''-1}, X^{p'-1})$ is the composition of the map $\pi^{*}: K^{n}(X^{p'-1},$ $X^{p-1}) \longrightarrow K^{n}(X^{p'-1})$ induced by $\pi: X^{p'-1} \rightarrow X^{p'-1} / X^{p-1}$, and the K-theory Bockstein map $\delta^{n}: K^{n}(X^{p'-1}) \longrightarrow K^{n+1}(X^{p''-1},X^{p'-1})$.
\end{itemize}

\subsection{K-theory and simplicial cohomology}

In the proof of the following lemma we will need the definition of \emph{reduced} and \emph{unreduced suspension} of a topological space $X$. We recall that the unreduced suspension is defined as $\hat{S}^{1}X = (X \times [-1,1]) / (X \times \{-1\}, X \times \{1\})$, i.e.\ as the double cone built on $X$. Instead, fixing a marked point $x_{0} \in X$, the reduced suspension is defined as $S^{1}X = \hat{S}^{1}X / (\{x_{0}\} \times [-1,1])$. The group $K^{1}(X)$ is defined as $K(S^{1}X)$, but, since $S^{1}X$ is obtained from $\hat{S}^{1}X$ quotienting out by a contractible subspace, it follows that $K(S^{1}X) \simeq K(\hat{S}^{1}X)$ \cite{Atiyah}.

\begin{Lemma}\label{OnePointUnion} For $k \in \mathbb{N}$ and $0 \leq i \leq k$, let:
	\[X = \dot{\bigcup_{i = 0, \ldots, k}} \; X_{i}
\]
be the one-point union of $k$ topological spaces. Then:
	\[\tilde{K}^{n}(X) \simeq \bigoplus_{i = 0}^{k} \tilde{K}^{n}(X_{i}) \; .
\]
\end{Lemma}
\textbf{Proof:} For $n = 0$, let us construct the isomorphism $\varphi: \tilde{K}(X) \rightarrow \bigoplus \tilde{K}(X_{i})$: it is simply given by $\varphi(\alpha)_{i} = \alpha\vert_{X_{i}}$, where $\alpha\vert_{X_{i}}$ is the pull-back via the immersion $X_{i} \rightarrow X$. To build $\varphi^{-1}$, let us consider $\{ [E_{i}] - [n_{i}] \} \in \bigoplus \tilde{K}(X_{i})$, where $[n_{i}]$ is the K-theory class represented by the trivial bundle of rank $n_{i}$. Since the sum is finite, by adding and subtracting a trivial bundle we can suppose $n_{i} = n_{j}$ for every $i,j$, so that we consider $\{ [E_{i}] - [n]\}$. Since the intersection of the $X_{i}$ is a point and the bundles $E_{i}$ have the same rank, we can glue them to a bundle $E$ on $X$ (see \cite{Atiyah} pp.\ 20-21): then we declare $\varphi^{-1} ( \, \{ [E_{i}] - [n] \} \, ) = ([E] - [n])$.

For $n = 1$, we first note that $\tilde{K}(\hat{S}^{1}(X_{1} \,\dot{\cup}\, X_{2})) = \tilde{K} (\hat{S}^{1}X_{1} \,\dot{\cup}\, \hat{S}^{1}X_{2})$, since quotienting by a contractible space (the linking between vertices of the cones and the joining point) we obtain the same space. Hence $\tilde{K}^{1}(X_{1} \,\dot{\cup}\, X_{2}) \simeq \tilde{K}^{1}(X_{1}) \oplus \tilde{K}^{1}(X_{2})$. Then, by induction, the thesis extends to finite families. Hence we have proven the result for $\tilde{K}^{n}$ with $n = 0$ and $n = 1$: by Bott periodicity \cite{Atiyah} the result holds for any $n$. $\square$

\paragraph{Remark:} we stress the fact that the previous lemma holds only for the one-point union of a \emph{finite} number of spaces.

\paragraph{}In the following theorem we suppose that the group of simplicial cochains $C^{p}(X, \mathbb{Z})$ of a finite simplicial complex coincides with the group of chains $C_{p}(X, \mathbb{Z})$: that's because, being the dimension finite, we can define the coboundary operator $\delta^{p}$ directly on chains, asking that the coboundary of a simplicial $p$-simplex $\sigma^{p}$ is the alternated sum of the $(p+1)$-simplices whose boundary contains $\sigma^{p}$ (while the boundary operator $\partial^{p}$ gives the alternated sum of the $(p-1)$-simplices contained in the boundary of $\sigma^{p}$). We can use this definition since the group of $p$-cochiains as usually defined, i.e.\ $\Hom(C_{p}(X, \mathbb{Z}), \mathbb{Z})$, is canonically isomorphic to $C_{p}(X, \mathbb{Z})$ in the case of finite simplicial complexes, and the usual coboundary operator corresponds to the one we defined above under such an isomorphism.

\begin{Theorem}\label{KTheoryCohomology} Let $X$ be a $n$-dimensional finite simplicial complex, $X^{p}$ be the $p$-skeleton of $X$ for $0 \leq p \leq n$ and $C^{p}(X, \mathbb{Z})$ be the group of simplicial $p$-cochains. Then, for any $p$ such that $0 \leq 2p \leq n$ or $0 \leq 2p+1 \leq n$, there are isomorphisms:
	\[\begin{split}
	&\Phi^{2p}: C^{2p}(X, \mathbb{Z}) \overset{\simeq}\longrightarrow K(X^{2p}, X^{2p-1})\\
	&\Phi^{2p+1}: C^{2p+1}(X, \mathbb{Z}) \overset{\simeq}\longrightarrow K^{1}(X^{2p+1}, X^{2p})
\end{split}\]
which can be summarized by:
	\[\Psi^{p}: C^{p}(X, \mathbb{Z}) \overset{\simeq}\longrightarrow K^{p}(X^{p}, X^{p-1}) \; .
\]
Moreover:
	\[K^{1}(X^{2p}, X^{2p-1}) = K(X^{2p+1}, X^{2p}) = 0 \; .
\]
\end{Theorem}
\paragraph{Proof:} We denote the simplicial structure of $X$ by $\Delta = \{\Delta^{m}_{i}\}$, where $m$ is the dimension of the simplex and $i$ enumerates the $m$-simplices, so that $X^{2p} = \displaystyle\bigcup_{i = 0}^{k} \Delta_{i}^{2p}$. Then the quotient by $X^{2p-1}$ is homeomorphic to $k$ spheres of dimension $2p$ attached to a point:
	\[X^{2p} / X^{2p-1} = \dot{\bigcup_{i}} \; S^{2p}_{i} \; .
\]
By lemma \ref{OnePointUnion} we obtain $\tilde{K}(X^{2p}/X^{2p-1}) \simeq \underset{i}\bigoplus \tilde{K}(S^{2p})$, and, by Bott periodicity, $\tilde{K}(S^{2p}) = \tilde{K}(S^{0}) = \mathbb{Z}$. Hence:
	\[K(X^{2p}, X^{2p-1}) \simeq \bigoplus_{i} \mathbb{Z} = C^{2p}(X, \mathbb{Z}) \; .
\]
For the odd case, let $X^{2p+1} = \displaystyle\bigcup_{j = 0}^{h} \Delta_{j}^{2p+1}$. We have by lemma \ref{OnePointUnion}:
	\[\begin{split}
	K^{1}(X^{2p+1}, X^{2p}) &= \tilde{K}^{1}\Bigl(\dot{\bigcup_{j}} \; S^{2p+1}_{j} \Bigr) = \bigoplus_{j} \tilde{K}^{1}\bigl(S^{2p+1}_{j}\bigr)\\
	&= \bigoplus_{j} \tilde{K}(S^{2p+2}_{j}) = \bigoplus_{j} \mathbb{Z} = C^{2p+1}(X, \mathbb{Z}) \; .
\end{split}\]
In the same way, $K^{1}(X^{2p}, X^{2p-1}) = \bigoplus_{j} \tilde{K}^{1}(S^{2p}_{j}) = \bigoplus_{j} \tilde{K}(S^{2p+1}_{j}) = 0$, and similarly for $K(X^{2p+1}, X^{2p})$. $\square$

\paragraph{}For $\eta$ the dual of the tautological line bundle on $\mathbb{CP}^{1}$, whose sheaf of sections is usually denoted as $\mathcal{O}_{\mathbb{CP}^{1}}(1)$, if we identify $S^{2}$ with $\mathbb{CP}^{1}$ topologically, the explicit isomorphisms $\Phi^{2p}$ and $\Phi^{2p+1}$ of theorem \ref{KTheoryCohomology} are:
	\[\Phi^{2p} \bigl(\Delta^{2p}_{i}\bigr) = \left\{\begin{matrix}
	(-1)^{p}(\eta - 1)^{\boxtimes p} & \in \tilde{K} \bigl(S^{2p}_{i}\bigr) &\\ \\
	0 & \in \tilde{K} \bigl(S^{2p}_{j}\bigr) & \text{ for $j \neq i$}
	\end{matrix}\right.
\]
and:
	\[\Phi^{2p+1} \bigl(\Delta^{2p+1}_{i}\bigr) = \left\{\begin{matrix}
	(-1)^{p+1}(\eta - 1)^{\boxtimes (p+1)} & \in \tilde{K}^{1} \bigl(S^{2p+1}_{i}\bigr) &\\ \\
	0 & \in \tilde{K}^{1} \bigl(S^{2p+1}_{j}\bigr) & \text{ for $j \neq i$}
	\end{matrix}\right.
\]
where we put the overall factors $(-1)^{p}$ and $(-1)^{p+1}$ for coherence with \eqref{LambdaR2n}.

\paragraph{Remark:} such isomorphisms are canonical, since every simplex is supposed to be oriented and $\eta - 1$ is distinguishable from $1 - \eta$ also up to automorphisms of $X$ (in the first case the trivial bundle has negative coefficient, in the second case the non-trivial one, so that, for example, they have opposite first Chern class).

\subsection{The spectral sequence}

We now recall how to build the spectral sequence. The assigned groups are:
	\[H^{n}(p,p') = K^{n}(X^{p'-1}, X^{p-1}) \; .
\]

\subsubsection{The first step}

The first step, from \eqref{Epr}, is:
	\[E^{p,\,q}_{1} = H^{p+q}(p, p+1) = K^{p+q}(X^{p}, X^{p-1}) \; .
\]
By theorem \ref{KTheoryCohomology} we have the isomorphisms:
	\[\begin{array}{lllllll}
	E^{2p, \,0}_{1} & \simeq & C^{2p}(X, \mathbb{Z}) & & E^{2p, \,1}_{1} & = & 0\\
	E^{2p+1, \,0}_{1} &\simeq & C^{2p+1}(X, \mathbb{Z}) & & E^{2p+1, \,1}_{1} & = & 0 \; .
\end{array}\]
Since, for a point $x_{0}$, $K(\{x_{0}\}) = \mathbb{Z}$ and $K^{1}(\{x_{0}\}) = 0$, we can write these isomorphisms in a compact form:
\begin{equation}\label{IsomCp}
	E^{p, \,q}_{1} \simeq C^{p}(X, K^{q}(x_{0})) \; .
\end{equation}
Anyway, since $E^{p, \,1}_{1} = 0$ for every $p$, and since only the parity of $q$ is meaningful, the only interesting case is $q = 0$. Therefore, from now on we deal only with the groups $E^{p, \,0}_{r}$. For the coboundaries, since $d^{p,\,q}_{r}: E^{p, \,q}_{r} \rightarrow E^{p+r, \,q-r+1}_{r}$, in particular $d^{p,\,0}_{r}: E^{p, \,0}_{r} \rightarrow E^{p+r, \,-r+1}_{r}$, if $r$ is even the coboundary is surely $0$, thus only the odd coboundaries are interesting. Therefore, from now on we deal only with the coboundaries $d^{p, \,0}_{r}$ with $r$ odd.

For $r = 1$, in the diagram \eqref{BoundaryDiagram} one has $\psi_{1} = \psi_{2} = \id$, hence $d^{p,\,0}_{1} = \delta_{2}$, i.e.\ $d^{p,\,0}_{1} = (\delta^{p})^{p,p+1,p+2}$. In particular:
	\[d^{p,\,0}_{1}: K^{p}(X^{p}, X^{p-1}) \longrightarrow K^{p+1}(X^{p+1}, X^{p})
\]
is the composition:
\begin{displaymath}
\xymatrix{
\tilde{K}^{p}(X^{p}/X^{p-1}) \ar[rr]^{d^{p, \,0}_{1}} \ar[dr]_{(\pi^{p,p-1})^{*}} & & \tilde{K}^{p+1}(X^{p+1}/X^{p})\\
& \tilde{K}^{p}(X^{p}). \ar[ur]_{\delta^{p}}
}
\end{displaymath}
for $\pi^{p,p-1}: X^{p} \rightarrow X^{p}/X^{p-1}$ the natural projection and $\delta^{p}$ is the Bockstein map. Another way to describe $d_{1}^{p,\,0}$ can be obtained considering the exact sequence induced by $X^{p}/X^{p-1} \longrightarrow X^{p+1}/X^{p-1} \longrightarrow X^{p+1}/X^{p}$: then $d_{1}^{p, \,0}$ is the corresponding Bockstein map:
\begin{equation}\label{d1p}
	d_{1}^{p, \,0}: \, \tilde{K}^{p}(X^{p}/X^{p-1}) \longrightarrow \tilde{K}^{p+1}(X^{p+1}/X^{p}) \; .
\end{equation}

\subsubsection{The second step}

We have shown that $E^{p, \,0}_{1} \simeq C^{p}(X, \mathbb{Z})$; we also have that $E^{p, \,0}_{2} \simeq H^{p}(X, \mathbb{Z})$ (see \cite{AH}), i.e.\ $d^{p, \,0}_{1}$ is the simplicial coboundary operator under the isomorphism \eqref{IsomCp}. By the first formula of \eqref{Epr} we have $E^{p,\,0}_{2} = \IIm \bigl( H^{p}(p, p+2) \overset{\psi^{p}}\longrightarrow H^{p}(p-1, p+1) \bigr)$, i.e.:
\begin{equation}\label{Ep2}
	E^{p, \,0}_{2} = \IIm \bigl( K^{p}(X^{p+1}, X^{p-1}) \overset{\psi^{p}}\longrightarrow K^{p}(X^{p}, X^{p-2}) \bigr) \; .
\end{equation}
Thus there is a canonical isomorphism:
\begin{equation}\label{IsomXi}
	\Xi^{p}: H^{p}(X, \mathbb{Z}) \longrightarrow \IIm \, \psi^{p} \subset K^{p}(X^{p}, X^{p-2}) \; .
\end{equation}

\paragraph{Cocycles and coboundaries} We now consider the maps:
	\[\begin{split}
	&\tilde{j}^{p}: X^{p}/X^{p-1} \longrightarrow X^{p+1}/X^{p-1}\\
	&\tilde{\pi}^{p}: X^{p}/X^{p-2} \longrightarrow X^{p}/X^{p-1} = \frac{X^{p}/X^{p-2}}{X^{p-1}/X^{p-2}}\\
	&\tilde{f}^{p}: X^{p}/X^{p-2} \longrightarrow X^{p+1}/X^{p-1}
\end{split}\]
These maps induce a commutative diagram:
\begin{equation}\label{DiagramE1E2}
\xymatrix{
E^{p, \,0}_{1} = \tilde{K}^{p}(X^{p}/X^{p-1}) \ar[dr]^{(\tilde{\pi}^{p})^{*}} & \\
\tilde{K}^{p}(X^{p+1}/X^{p-1}) \ar[u]^{(\tilde{j}^{p})^{*}} \ar[r]^{(\tilde{f}^{p})^{*}} & \tilde{K}^{p}(X^{p}/X^{p-2})
}
\end{equation}
where $(\tilde{f}^{p})^{*}, (\tilde{j}^{p})^{*}, (\tilde{\pi}^{p})^{*}$ are maps of the $\psi$-type. We have that $E^{p, \,0}_{2} = \IIm (\tilde{f}^{p})^{*}$ by \eqref{Ep2}. We now prove that:
\begin{enumerate}
	\item $\Ker \, d^{p, \,0}_{1} = \IIm (\tilde{j}^{p})^{*}$;
	\item $\IIm \, d^{p-1, \, 0}_{1} = \Ker (\tilde{\pi}^{p})^{*}$.
\end{enumerate}
The first statement follows directly from \eqref{d1p} using the exact sequence:
	\[\cdots \longrightarrow \tilde{K}^{p}(X^{p+1}/X^{p-1}) \overset{(\tilde{j}^{p})^{*}}\longrightarrow \tilde{K}^{p}(X^{p}/X^{p-1}) \overset{d^{p, \,0}_{1}}\longrightarrow \tilde{K}^{p+1} (X^{p+1}/X^{p}) \longrightarrow \cdots
\]
and the second by the exact sequence:
	\[\cdots \longrightarrow \tilde{K}^{p-1}(X^{p-1}/X^{p-2}) \overset{d^{p-1, \, 0}_{1}}\longrightarrow \tilde{K}^{p}(X^{p}/X^{p-1}) \overset{(\tilde{\pi}^{p})^{*}}\longrightarrow \tilde{K}^{p}(X^{p}/X^{p-2}) \longrightarrow \cdots.
\]
Since $\IIm (\tilde{f}^{p})^{*} \simeq H^{p}(X, \mathbb{Z})$ and $d^{p,\,0}_{1}$ corresponds to the simplicial coboundary under this isomorphism, it follows that:
\begin{itemize}
	\item cocycles in $C^{p}(X, \mathbb{Z})$ correspond to classes in $\IIm (\tilde{j}^{p})^{*}$, i.e.\ to classes in $\tilde{K}^{p}(X^{p}/X^{p-1})$ that are restriction of classes in $\tilde{K}^{p}(X^{p+1}/X^{p-1})$;
	\item coboundaries in $C^{p}(X, \mathbb{Z})$ corresponds to classes in $\Ker (\tilde{\pi}^{p})^{*}$, i.e.\ to classes in $\tilde{K}^{p}(X^{p}/X^{p-1})$ that are $0$ when lifted to $\tilde{K}^{p}(X^{p}/X^{p-2})$;
	\item $\IIm \, \pi^{*}$ corresponds to cochains (not only cocycles) up to coboundaries and its subset $\IIm (\tilde{f}^{p})^{*}$ corresponds to cohomology classes;
	\item given $\alpha \in \IIm (\tilde{f}^{p})^{*}$, we can lift it to a class in $\tilde{K}^{p}(X^{p} / X^{p-1})$ choosing different trivializations on $X^{p-1}/X^{p-2}$, and the different homotopy classes of such trivializations determine the different respresentative cocycles of the class.
\end{itemize}

\subsubsection{The last step}

We recall equation \eqref{Limit}:
	\[E^{p,\,q}_{\infty} = \IIm \bigl( H^{p+q}(p, +\infty) \overset{\psi^{p+q}}\longrightarrow H^{p+q}(0, p+1) \bigr)
\]
which, in our case, becomes:
\begin{equation}\label{EpinftyA}
	E^{p, \,0}_{\infty} = \IIm \bigl( K^{p}(X, X^{p-1}) \overset{\psi^{p}}\longrightarrow K^{p}(X^{p}) \bigr)
\end{equation}
where $\psi$ is obtained by the pull-back of $f^{p}: X^{p} \rightarrow X / X^{p-1}$. Since, for $i^{p}: X^{p} \rightarrow X$ the natural immersion and $\pi^{p} : X \rightarrow X/X^{p}$ the natural projection, $f^{p} = \pi^{p-1} \circ i^{p}$ holds, the following diagram commutes:
\begin{equation}\label{EpinftyB}
\xymatrix{
\tilde{K}^{p}(X/X^{p-1}) \ar[dr]_{(\pi^{p-1})^{*}} \ar[rr]^{(f^{p})^{*}} & & \tilde{K}^{p}(X^{p})\\
& \tilde{K}^{p}(X) \ar[ur]_{(i^{p})^{*}}.
}
\end{equation}

\paragraph{Remark:} in the previous triangle we cannot say that $(i^{p})^{*} \circ (\pi^{p-1})^{*} = 0$ by exactness, since by exactness $(i^{p})^{*} \circ (\pi^{p})^{*} = 0$  at the same level $p$, as follows from $X^{p} \rightarrow X \rightarrow X/X^{p}$.

\paragraph{}The sequence $K^{p}(X, X^{p-1}) \overset{(\pi^{p-1})^{*}}\longrightarrow K^{p}(X) \overset{(i^{p-1})^{*}}\longrightarrow K^{p}(X^{p-1})$ is exact, i.e.:
	\[\IIm (\pi^{p-1})^{*} = \Ker (i^{p-1})^{*} \; .
\]
Since trivially $\Ker (i^{p})^{*} \subset \Ker (i^{p-1})^{*}$, we obtain that $\Ker (i^{p})^{*} \subset \IIm (\pi^{p-1})^{*}$. Moreover:
	\[\IIm \, \psi = \IIm \, \bigl( (i^{p})^{*} \circ (\pi^{p-1})^{*} \bigr) = \IIm \, \Bigl( (i^{p})^{*} \; \big\vert_{\IIm (\pi^{p-1})^{*}} \Bigr) \simeq \frac{\, \IIm (\pi^{p-1})^{*} \,} {\Ker (i^{p})^{*}} = \frac{\, \Ker \, (i^{p-1})^{*} \,} {\Ker (i^{p})^{*}}
\]
hence, finally:
\begin{equation}
	E^{p, \,0}_{\infty} \simeq \frac{\, \Ker \bigl( K^{p}(X) \longrightarrow K^{p}(X^{p-1}) \bigr) \,} {\Ker \bigl( K^{p}(X) \longrightarrow K^{p}(X^{p}) \bigr)}\\
\end{equation}
i.e.\ $E^{p, \, 0}_{\infty}$ is made, up to canonical isomorphism, by $p$-classes on $X$ which are $0$ on $X^{p-1}$, up to classes which are $0$ on $X^{p}$.

\subsubsection{From the first to the last step}

We now see how to link the first and the last step of the sequence. In general we have:
	\[E^{p,\,q}_{1} = H^{p+q}(p, p+1) \qquad E^{p,\,q}_{\infty} = \IIm \bigl( H^{p+q}(p, +\infty) \overset{\psi_{1}}\longrightarrow H^{p+q}(0, p+1) \bigr) \; .
\]
for $\psi_{1} = (\psi^{p+q})^{p, +\infty}_{0, p+1}$. We also consider the map:
	\[\psi_{2}: H^{p+q}(p, +\infty) \longrightarrow H^{p+q}(p, p+1)
\]
where $\psi_{2} = (\psi^{p+q})^{p, +\infty}_{p, p+1}$. An element $\alpha \in E^{p,\,q}_{1}$ survives until the last step if and only if $\alpha \in \IIm \, \psi_{2}$ and its class in $E^{p,\,q}_{\infty}$ is $\psi_{1} \circ (\psi_{2}^{-1})(\alpha)$, which is well-defined since $\Ker \, \psi_{2} \subset \Ker \, \psi_{1}$. For:
	\[\psi_{3}: H^{p+q}(p, p+1) \longrightarrow H^{p+q}(0, p+1)
\]
i.e.\ $\psi_{3} = (\psi^{p+q})^{p, p+1}_{0, p+1}$, it holds that $\psi_{1} = \psi_{3} \circ \psi_{2}$, so that $\psi_{1} \circ (\psi_{2}^{-1}) = \psi_{3}$. For $\alpha \in \IIm \, \psi_{2} \subset E^{p,\,q}_{1}$, we call $\{\alpha\}_{E^{p,\,q}_{\infty}}$ the class it reaches in $E^{p,\,q}_{\infty}$. Then we have:
	\[\{\alpha\}_{E^{p,\,q}_{\infty}} = \psi_{3}(\alpha) \; .
\]

\paragraph{}For AHSS this becomes:
	\[E^{p, \,0}_{1} = K^{p}(X^{p}, X^{p-1}) \qquad E^{p, \,0}_{\infty} = \IIm \bigl( K^{p}(X, X^{p-1}) \overset{\psi_{1}}\longrightarrow K^{p}(X^{p}) \bigr)
\]
and:
	\[\psi_{2}: K^{p}(X, X^{p-1}) \longrightarrow K^{p}(X^{p}, X^{p-1}) \; .
\]
In this case, $\psi_{2} = (i^{p,p-1})^{*}$ for $i^{p,p-1}: X^{p} / X^{p-1} \rightarrow X / X^{p-1}$. Thus, the classes in $E^{p, \,0}_{1}$ surviving until the last step are the ones which are restrictions of a class defined on all $X / X^{p-1}$. Moreover, $\psi_{1} = (f^{p})^{*}$ for $f^{p}: X^{p} \rightarrow X / X^{p-1}$, and $f^{p} = i^{p,p-1} \circ \pi^{p,p-1}$ for $\pi^{p,p-1}: X^{p} \rightarrow X^{p} / X^{p-1}$. Hence $\psi_{1} = (\pi^{p,p-1})^{*} \circ \psi_{2}$, and, in fact, $\psi_{3} = (\pi^{p,p-1})^{*}$. This implies that, for $\alpha \in \IIm \, \psi_{2} \subset E^{p, \,0}_{1}$:
\begin{equation}\label{FromOneToInfty}
	\{\alpha\}_{E^{p, \,0}_{\infty}} = (\pi^{p,p-1})^{*}(\alpha) \; .
\end{equation}

\subsection{Rational K-theory and cohomology}

We now consider the Atiyah-Hirzebruch spectral sequence in the rational case \cite{AH}. In particular, we consider the groups:
	\[H^{n}(p,p') = K_{\mathbb{Q}}^{n}(X^{p'-1}, X^{p-1})
\]
where $K_{\mathbb{Q}}^{n}(X, Y) := K^{n}(X, Y) \otimes_{\mathbb{Z}} \mathbb{Q}$. In this case the sequence is made by the groups $Q^{p, \, q}_{r} = E^{p, \, q}_{r} \otimes_{\mathbb{Z}} \,\mathbb{Q}$. In particular:
\begin{equation}\label{QSequence}
\begin{array}{lll}
		Q^{p, \, 0}_{1} \simeq C^{p}(X, \mathbb{Q}) & & Q^{p, \, 1}_{1} = 0\\ \\
		Q^{p, \, 0}_{2} \simeq H^{p}(X, \mathbb{Q}) & & Q^{p, \, 1}_{2} = 0\\ \\
		Q^{p, \, 0}_{\infty} \simeq \displaystyle\frac{\, \Ker \bigl( K_{\mathbb{Q}}^{p}(X) \longrightarrow \displaystyle K_{\mathbb{Q}}^{p}(X^{p-1}) \bigr) \,} {\Ker \bigl( K_{\mathbb{Q}}^{p}(X) \longrightarrow \displaystyle K_{\mathbb{Q}}^{p}(X^{p}) \bigr)} & & Q^{p, \, 1}_{\infty} = 0 \; .
	\end{array}
\end{equation}
Such a sequence collapses at the second step \cite{AH}, hence $Q^{p, \, 0}_{\infty} \simeq Q^{p, \, 0}_{2}$. Since:
\begin{itemize}
	\item $\bigoplus_{p} Q^{p, \, 0}_{\infty}$ is the graded group associated to the chosen filtration of $K(X) \oplus K^{1}(X)$;
	\item in particular, by \eqref{QSequence}, $\bigoplus_{2p} Q^{2p, \, 0}_{\infty}$ is the graded group of $K_{\mathbb{Q}}(X)$ and $\bigoplus_{2p+1} Q^{2p+1, \, 0}_{\infty}$ is the graded group of $K^{1}_{\mathbb{Q}}(X)$;
	\item $Q^{p, \, 0}_{\infty} \simeq H^{p}(X, \mathbb{Q})$, thus it has no torsion;
\end{itemize}
it follows that:
	\[K_{\mathbb{Q}}(X) = \bigoplus_{2p} Q^{2p, \, 0}_{\infty} \qquad K_{\mathbb{Q}}^{1}(X) = \bigoplus_{2p+1} Q^{2p+1, \, 0}_{\infty}
\]
hence:
	\[K_{\mathbb{Q}}(X) \simeq H^{\ev}(X, \mathbb{Q}) \qquad K^{1}_{\mathbb{Q}}(X) \simeq H^{\odd}(X, \mathbb{Q}).
\]
In particular, the isomorphisms of the last equation are given by the Chern character:
	\[\begin{split}
	&\ch: K_{\mathbb{Q}}(X) \longrightarrow H^{\ev}(X, \mathbb{Q})\\
	&\ch: K^{1}_{\mathbb{Q}}(X) \longrightarrow H^{\ev}(S^{1}X, \mathbb{Q}) \simeq H^{\odd}(X, \mathbb{Q})
\end{split}\]
and they are also isomorphism of rings.

\section{Gysin map and AHSS}

We are now ready to describe the explicit link between the Gysin map and the Atiyah-Hirzebruch spectral sequence. We start with the case of an embedded sumbanifold of even codimension, corresponding, from a physical point of view, to a D-brane world-volume in type IIB superstring theory, then we reproduce the same result in the case of odd codimension, corresponding to a D-brane world-volume in type IIA superstring theory.

\subsection{Even case}

We call $X$ a compact smooth $n$-dimensional manifold and $Y$ a compact embedded $p$-dimensional submanifold. We choose a finite triangulation of $X$ which restricts to a triangulation of $Y$ \cite{Munkres}. We use the following notation:
\begin{itemize}
	\item we denote the triangulation of $X$ by $\Delta = \{\Delta^{m}_{i}\}$, where $m$ is the dimension of the simplex and $i$ enumerates the $m$-simplices;
	\item we denote by $X_{\Delta}^{p}$ the $p$-skeleton of $X$ with respect to $\Delta$.
\end{itemize}
The same notation is used for other triangulations or simplicial decompositions of $X$ and $Y$. In the following theorem we need the definition of ``dual cell decomposition'' with respect to a triangulation: we refer to \cite{GH} pp.\ 53-54.
\begin{Theorem}\label{Triangulation} Let $X$ be an $n$-dimensional compact manifold and $Y \subset X$ a $p$-dimensional embedded compact submanifold. Let:
\begin{itemize}
	\item $\Delta = \{\Delta^{m}_{i}\}$ be a triangulation of $X$ which restricts to a triangulation $\Delta' = \{\Delta^{m}_{i'}\}$ of $Y$;
	\item $D = \{D^{n-m}_{i}\}$ be the dual decomposition of $X$ with respect to $\Delta$;
	\item $\tilde{D} \subset D$ be subset of $D$ made by the duals of the simplices in $\Delta'$.
\end{itemize}
Then, calling $\abs{\tilde{D}}$ the support of $\tilde{D}$:
\begin{itemize}
	\item the interior of $\abs{\tilde{D}}$ is a tubular neighborhood of $Y$ in $X$;
	\item the interior of $\abs{\tilde{D}}$ does not intersect $X_{D}^{n-p-1}$, i.e.:
	\[\abs{\tilde{D}} \cap X_{D}^{n-p-1} \subset \partial \abs{\tilde{D}} \; .
\]
\end{itemize}
\end{Theorem}
\textbf{Proof:} The $n$-simplices of $\tilde{D}$ are the duals of the vertices of $\Delta'$. Let $\tau = \{\tau^{m}_{j}\}$ be the first baricentric subdivision of $\Delta$ \cite{GH,Hatcher}. For each vertex $\Delta^{0}_{i'}$ in $Y$ (thought of as an element of $\Delta$), its dual is:
\begin{equation}\label{DTilde}
	\tilde{D}^{n}_{i'} = \bigcup_{\Delta^{0}_{i'} \in \tau^{n}_{j}} \tau^{n}_{j} \; .
\end{equation}
Moreover, if $\tau' = \{{\tau'}^{m}_{j'}\}$ is the first baricentric subdivision of $\Delta'$ (of course $\tau' \subset \tau$) and $D' = \{{D'}^{m}_{i'}\}$ is the dual of $\Delta'$ in $Y$, then (reminding that $p$ is the dimension of $Y$):
\begin{equation}\label{DPrimo}
	D'^{\,p}_{\;\,i'} = \bigcup_{\Delta^{0}_{i'} \in {\tau'}^{p}_{j'}} {\tau'}^{p}_{j'}
\end{equation}
and:
	\[\tilde{D}^{n}_{i'} \cap Y = D'^{\,p}_{\;\,i'} \; .
\]
Moreover, let us consider the $(n-p)$-simplices in $\tilde{D}$ contained in $\partial \tilde{D}^{n}_{i'}$ (for a fixed $i'$ in formula \eqref{DTilde}), i.e.\ $X^{n-p}_{\tilde{D}} \cap \tilde{D}^{n}_{i'}$: they intersect $Y$ transversally in the baricenters of each $p$-simplex of $\Delta'$ containing $\Delta^{0}_{i'}$: we call such baricenters $\{b_{1}, \ldots, b_{k}\}$ and the intersecting $(n-p)$-cells $\{\tilde{D}^{n-p}_{l}\}_{l = 1, \ldots, k}$. Since (for a fixed $i'$) $\tilde{D}^{n}_{i'}$ retracts on $\Delta^{0}_{i'}$, we can consider a local chart $(U_{i'}, \varphi_{i'})$, with $U_{i'} \subset \mathbb{R}^{n}$ neighborhood of $0$, such that:
\begin{itemize}
	\item $\varphi_{i'}^{-1}(U_{i'})$ is a neighborhood of $\tilde{D}^{n}_{i'}$;
	\item $\varphi_{i'}(D'^{\,p}_{\;\,i'}) \subset U_{i'} \cap (\{0\} \times \mathbb{R}^{p})$, for $0 \in \mathbb{R}^{n-p}$ (see eq. \eqref{DPrimo});
	\item $\varphi_{i'}(\tilde{D}^{n-p}_{l}) \subset U_{i'} \cap \bigl(\mathbb{R}^{n-p} \times \pi_{p}(\varphi_{i'}(b_{l}))\bigr)$, for $\pi_{p}: \mathbb{R}^{n} \rightarrow \{0\} \times \mathbb{R}^{p}$ the projection.
\end{itemize}
We now consider the natural foliation of $U_{i'}$ given by the intersection with the hyperplanes $\mathbb{R}^{n-p} \times \{x\}$ and its image via $\varphi_{i'}^{-1}$: in this way, we obain a foliation of $\tilde{D}^{n}_{i'}$ transversal to $Y$. If we do this for any $i'$, by construction the various foliations glue on the intersections, since such intersections are given by the $(n-p)$-cells $\{\tilde{D}^{n-p}_{l}\}_{l = 1, \ldots, k}$, and the interior gives a $C^{0}$-tubular neighborhood of $Y$.

Moreover, a $(n-p-r)$-cell of $\tilde{D}$, for $r > 0$, cannot intersect $Y$ since it is contained in the boundary of a $(n-p)$-cell, and such cells intersect $Y$, which is done by $p$-cells, only in their interior points $b_{j}$. Being the simplicial decomposition finite, it follows that the interior of $\abs{\tilde{D}}$ does not intersect $X_{D}^{n-p-1}$. \\
$\square$

\paragraph{}We now consider quintuples $(X, Y, \Delta, D, \tilde{D})$ satisfying the following condition:
\begin{itemize}
	\item[$(\#)$] $X$ is an $n$-dimensional compact manifold and $Y \subset X$ a $p$-dimensional embedded compact submanifold, such that $n - p$ is even and the normal bundle $\mathcal{N}(Y)$ is spin$^{c}$. Moreover, $\Delta$, $D$ and $\tilde{D}$ are defined as in theorem \ref{Triangulation}.
\end{itemize}

\begin{Lemma}\label{TrivialityXnp1} Let $(X,Y,\Delta,D,\tilde{D})$ be a quintuple satisfying $(\#)$, $U = \Int\abs{\tilde{D}}$ and $\alpha \in K(Y)$. Then:
\begin{itemize}
	\item there exists a neighborhood $V$ of $X \setminus U$ such that $i_{!}(\alpha) \vert_{V} = 0$;
	\item in particular, $i_{!}(\alpha) \,\vert_{X^{n-p-1}_{D}} = 0$.
\end{itemize}
\end{Lemma}
\textbf{Proof:} By equation \eqref{GysinMap}:
	\[i_{!}(\alpha) = \psi^{*} \beta, \qquad \beta = (\varphi_{U}^{+})^{*} \circ T(\alpha) \in \tilde{K}(U^{+}) \; .
\]
Let $\beta = [E] - [n]$, and let $V_{\infty} \subset U^{+}$ be a neighborhood of $\infty$ which trivializes $E$. Then $(\psi^{*}E) \,\big\vert_{\psi^{-1}(V_{\infty})}$ is trivial. Hence, for $V = \psi^{-1}(V_{\infty})$:
	\[(\psi^{*} \beta)\big\vert_{V} = \bigl[(\psi^{*}E)\big\vert_{V}\bigr] - [n] = [n] - [n] = 0 \; .
\]
By theorem \ref{Triangulation}, $X^{n-p-1}_{D}$ does not intersect the tubular neighborhood $\Int \abs{\tilde{D}}$ of $Y$, hence $X^{n-p-1}_{D} \subset \psi^{-1}(V_{\infty}) = V$, so that $(\psi^{*} \beta)\big\vert_{X^{n-p-1}_{D}} = 0$. $\square$

\subsubsection{Trivial bundle}

We start considering the case of a trivial bundle.

\begin{Theorem}\label{FirstTheorem} Let $(X,Y,\Delta,D,\tilde{D})$ be a quintuple satisfying $(\#)$ and $\Phi^{n-p}_{D}: C^{n-p}(X, \mathbb{Z}) \rightarrow K(X_{D}^{n-p},$ $X_{D}^{n-p-1})$ be the isomorphism stated in theorem \ref{KTheoryCohomology}. Let:
	\[\pi^{n-p,\,n-p-1}: X_{D}^{n-p} \longrightarrow X_{D}^{n-p} / X_{D}^{n-p-1}
\]
be the projection and $\PD_{\Delta}Y$ be the representative of $\PD_{X}[Y]$ (for $[Y]$ the homology class of $Y$) given by the sum of the cells dual to the $p$-cells of $\Delta$ covering $Y$. Then:
	\[i_{!}(Y \times \mathbb{C}) \vert_{X_{D}^{n-p}} = (\pi^{n-p,\,n-p-1})^{*}( \Phi_{D}^{n-p}(\PD_{\Delta}Y)) \; .
\]
\end{Theorem}
\textbf{Proof:} We define:
	\[(U^{+})^{n-p}_{D} = \frac{\overline{X^{n-p}_{D} \, \vert_{U}}}{X^{n-p-1}_{D} \, \vert_{\partial U}}
\]
so that there is a natural immersion $(U^{+})^{n-p}_{D} \subset U^{+}$ defined sending the denominator to $\infty$ (the numerator is exactly $X^{n-p}_{\tilde{D}}$ of theorem \ref{Triangulation}). We also define, considering the map $\psi$ of equation \eqref{GysinMap}:
	\[\psi^{n-p} = \psi\big\vert_{X^{n-p}_{D}}: \, X^{n-p}_{D} \longrightarrow (U^{+})^{n-p}_{D} \; .
\]
The latter is well-defined since the $(n-p)$-simplices outside $U$ and all the $(n-p-1)$-simplices are sent to $\infty$ by $\psi$. Calling $I$ the set of indices of the $(n-p)$-simplices in $D$, calling $S^{k}$ the $k$-dimensional sphere and denoting by $\dot{\cup}$ the one-point union of topological spaces, there are the following canonical homeomorphisms:
	\[\begin{split}
	& \xi^{n-p}_{X}: \pi^{n-p}(X_{D}^{n-p}) \overset{\simeq}\longrightarrow \dot{\bigcup_{i \in I}} \; S^{n-p}_{i} \\
	& \xi^{n-p}_{U^{+}}: \psi^{n-p}(X_{D}^{n-p}) \overset{\simeq}\longrightarrow \dot{\bigcup_{j \in J}} \; S^{n-p}_{j}
\end{split}\]
where $\{S^{n-p}_{j}\}_{j \in J}$, with $J \subset I$, is the set of $(n-p)$-spheres corresponding to the $(n-p)$-simplices with interior contanined in $U$, i.e.\ corresponding to $\pi^{n-p}\bigl(\overline{X^{n-p}_{D} \, \big\vert_{U}}\,\bigr)$. The homeomorphism $\xi^{n-p}_{U^{+}}$ is due to the fact that the boundary of the $(n-p)$-cells intersecting $U$ is contained in $\partial U$, hence it is sent to $\infty$ by $\psi^{n-p}$, while all the $(n-p)$-cells outside $U$ are sent to $\infty$: hence, the image of $\psi^{n-p}$ is homeomorphic to $\dot{\bigcup}_{j \in J} \; S^{n-p}_{j}$ sending $\infty$ to the attachment point. We define:
	\[\begin{split}
	\rho: \, \dot{\bigcup_{i \in I}} \; S^{n-p}_{i} \longrightarrow \dot{\bigcup_{j \in J}} \; S^{n-p}_{j}
\end{split}\]
as the natural projection, i.e.\ $\rho$ is the identity of $S^{n-p}_{j}$ for every $j \in J$ and sends all the spheres in $\{S^{n-p}_{i}\,\}_{i \in I \setminus J}$ to the attachment point. We have that:
	\[\xi^{n-p}_{U^{+}} \circ \psi^{n-p} = \rho \circ \xi^{n-p}_{X} \circ \pi^{n-p,\,n-p-1}
\]
hence:
\begin{equation}\label{PsiPiRho}
	(\psi^{n-p})^{*} \circ (\xi^{n-p}_{U^{+}})^{*} = (\pi^{n-p,\,n-p-1})^{*} \circ (\xi^{n-p}_{X})^{*} \circ \rho^{*} \; .
\end{equation}
We put $\mathcal{N} = \mathcal{N}(Y)$ and $\tilde{\lambda}_{\mathcal{N}} = (\varphi_{U}^{+})^{*} (\lambda_{\mathcal{N}})$, where $\lambda_{\mathcal{N}}$ is the Thom class of the normal bundle defined in equation \eqref{ThomClass}. By lemma \ref{Unitarity} and equation \eqref{GysinMap} we have $i_{!}(Y \times \mathbb{C}) = \psi^{*} \circ (\varphi_{U}^{+})^{*} (\lambda_{\mathcal{N}})$. Then:
	\[i_{!}(Y \times \mathbb{C}) \, \big\vert_{X_{D}^{n-p}} = \psi^{*} (\tilde{\lambda}_{\mathcal{N}}) \, \big\vert_{X_{D}^{n-p}} = (\psi^{n-p})^{*} \bigl( \tilde{\lambda}_{\mathcal{N}} \,\big\vert_{(U^{+})^{n-p}_{D}} \bigr)
\]
and
	\[(\xi^{n-p}_{X})^{*} \circ \rho^{*} \circ ((\xi^{n-p}_{U^{+}})^{-1})^{*} \bigl(\, \tilde{\lambda}_{\mathcal{N}} \,\big\vert_{(U^{+})^{n-p}_{D}} \,\bigr) = \Phi_{D}^{n-p}(\PD_{\Delta}Y)
\]
since:
\begin{itemize}
	\item $\PD_{\Delta}Y$ is the sum of the $(n-p)$-cells intersecting $U$;
	\item hence $((\xi^{n-p}_{X})^{-1})^{*} \circ \Phi_{D}^{n-p}(\PD_{\Delta}Y)$ gives a $(-1)^{\frac{n-p}{2}} (\eta - 1)^{\boxtimes \frac{n-p}{2}}$ factor to each sphere $S^{n-p}_{j}$ for $j \in J$ and $0$ otherwise;
	\item but this is exactly $\rho^{*} \circ ((\xi^{n-p}_{U^{+}})^{-1})^{*} \bigl( \tilde{\lambda}_{\mathcal{N}} \,\big\vert_{(U^{+})^{n-p}_{D}} \bigr)$ since by equation \eqref{RestrictionThom} we have, for $y \in Y$:
	\[(\lambda_{\mathcal{N}})\, \big\vert_{\mathcal{N}_{y}^{+}} = \lambda_{\mathbb{R}^{n-p}} \simeq (-1)^{\frac{n-p}{2}}(\eta - 1)^{\boxtimes \frac{n-p}{2}}
\]
and for the spheres outside $U$, that $\rho$ sends to $\infty$, we have that:
	\[\begin{split}
	\rho^{*} \bigl( \tilde{\lambda}_{\mathcal{N}} \,\big\vert_{(U^{+})^{n-p}_{D}} \bigr) \Big\vert_{\dot{\bigcup}_{i \in I \setminus J} \; S^{n-p}_{i}}
	&= \rho^{*} \bigl( \tilde{\lambda}_{\mathcal{N}} \,\big\vert_{\rho(\dot{\bigcup}_{i \in I \setminus J} \; S^{n-p}_{i})} \bigr)\\
	&= \rho^{*} \bigl( \tilde{\lambda}_{\mathcal{N}} \,\big\vert_{\{\infty\}} \bigr) = \rho^{*}(0) = 0 \; .
\end{split}\]
\end{itemize}
Hence, from equation \eqref{PsiPiRho}:
	\[\begin{split}
	i_{!}(Y \times \mathbb{C}) \, \big\vert_{X_{D}^{n-p}} &= (\psi^{n-p})^{*} \bigl( \tilde{\lambda}_{\mathcal{N}} \,\big\vert_{(U^{+})^{n-p}_{D}} \bigr)\\
	&= (\pi^{n-p,\,n-p-1})^{*} \circ (\xi^{n-p}_{X})^{*} \circ \rho^{*} \circ ((\xi^{n-p}_{U^{+}})^{-1})^{*}\bigl( \tilde{\lambda}_{\mathcal{N}} \,\big\vert_{(U^{+})^{n-p}_{D}} \bigr)\\
	&= (\pi^{n-p,\,n-p-1})^{*} \Phi_{D}^{n-p}(\PD_{\Delta}Y) \; .
\end{split}\]
$\square$

\paragraph{}The following theorem encodes the link between the Gysin map and the AHSS.

\begin{Theorem}\label{SecondTheorem} Let $(X,Y,\Delta,D,\tilde{D})$ be a quintuple satisfying $(\#)$ and $\Phi^{n-p}_{D}: C^{n-p}(X, \mathbb{Z}) \rightarrow K(X_{D}^{n-p}, X_{D}^{n-p-1})$ be the isomorphism stated in theorem \ref{KTheoryCohomology}. Let us suppose that $\PD_{\Delta}Y$ is contained in the kernel of all the boundaries $d^{n-p, \,0}_{r}$ for $r \geq 1$. Then it defines a class:
	\[\{\Phi^{n-p}_{D}(\PD_{\Delta}Y)\}_{E^{n-p, \,0}_{\infty}} \in E^{n-p,\,0}_{\infty} \simeq \frac{\, \Ker ( K(X) \longrightarrow K(X^{n-p-1}) ) \,} {\Ker ( K(X) \longrightarrow K(X^{n-p}) )} \; .
\]
The following equality holds:
	\[\{\Phi^{n-p}_{D}(\PD_{\Delta}Y)\}_{E^{n-p, \,0}_{\infty}} = [i_{!}(Y \times \mathbb{C})] \; .
\]
\end{Theorem}
\textbf{Proof:} By equations \eqref{EpinftyA} and \eqref{EpinftyB} we have the following commutative diagram:
\begin{equation}\label{Epinfty}
\xymatrix{
	E^{n-p, \,0}_{\infty} = \IIm \bigl( \tilde{K}(X/X_{D}^{n-p-1}) \ar[dr]_{(\pi^{n-p-1})^{*}} \ar[rr]^{\qquad\qquad (f^{n-p})^{*}} & & \tilde{K}(X_{D}^{n-p}) \bigr)\\
& \tilde{K}(X) \ar[ur]_{(i^{n-p})^{*}}
}
\end{equation}
and, given a representative $\alpha \in \IIm (\pi^{n-p-1})^{*} = \Ker ( \tilde{K}(X) \rightarrow \tilde{K}(X_{D}^{n-p-1}) )$, we have that $\{\alpha\}_{E^{n-p, \,0}_{\infty}} = (i^{n-p})^{*}(\alpha) = \alpha\vert_{X_{D}^{n-p}}$. Moreover:
\begin{itemize}
	\item the class $\{\Phi^{n-p}_{D}(\PD_{\Delta}Y)\}_{E^{n-p, \,0}_{\infty}}$, by formula \eqref{FromOneToInfty}, corresponds to the element of $\tilde{K}(X_{D}^{n-p})$ defined by $(\pi^{n-p,\,n-p-1})^{*}(\Phi^{n-p}_{D}(\PD_{\Delta}Y))$, for $\pi^{n-p,\,n-p-1}: X^{n-p}_{D} \rightarrow X^{n-p}_{D} / X^{n-p-1}_{D}$;
	\item by lemma \ref{TrivialityXnp1} we have $i_{!}(Y \times \mathbb{C}) \in \Ker(K(X) \longrightarrow K(X_{D}^{n-p-1}))$, hence $[i_{!}(Y \times \mathbb{C})]$ is well-defined as an element of $E^{n-p, \,0}_{\infty}$ and, by exactness, $i_{!}(Y \times \mathbb{C}) \in \IIm (\pi^{n-p-1})^{*}$;
	\item by theorem \ref{FirstTheorem} we have $(i^{n-p})^{*}(i_{!}(Y \times \mathbb{C})) = (\pi^{n-p,\,n-p-1})^{*}(\Phi_{D}^{n-p}(\PD(Y)))$;
	\item hence $\{\Phi^{n-p}_{D}(\PD_{\Delta}Y)\}_{E^{n-p, \,0}_{\infty}} = [i_{!}(Y \times \mathbb{C})]$.
\end{itemize}
$\square$

\paragraph{}Let us consider a trivial vector bundle of generic rank $Y \times \mathbb{C}^{r}$. We denote by $[r]$ its K-theory class on $Y$. By lemma \ref{Unitarity} we have that $[r] \cdot \lambda_{\mathcal{N}} = \lambda_{\mathcal{N}}^{\oplus r}$, hence theorem \ref{FirstTheorem} becomes:
	\[i_{!}(Y \times \mathbb{C}^{r}) \, \big\vert_{X_{D}^{n-p}} = (\pi^{n-p,\,n-p-1})^{*} \bigl( \Phi_{D}^{n-p}(\PD_{\Delta}(r \cdot Y)) \bigr)
\]
and theorem \ref{SecondTheorem} becomes:
	\[\{\Phi^{n-p}_{D}(\PD_{\Delta}(r \cdot Y))\}_{E^{n-p, \,0}_{\infty}} = [i_{!}(Y \times \mathbb{C}^{r})] \; .
\]

\subsubsection{Generic bundle}

If we consider a generic bundle $E$ over $Y$ of rank $r$, we can prove that $i_{!}(E)$ and $i_{!}(Y \times \mathbb{C}^{r})$ have the same restriction to $X^{n-p}_{D}$: in fact, the Thom isomorphism gives $T(E) = E \cdot \lambda_{\mathcal{N}}$ and, if we restrict $E \cdot \lambda_{\mathcal{N}}$ to a finite family of fibers, which are transversal to $Y$, the contribution of $E$ becomes trivial, so it has the same effect of the trivial bundle $Y \times \mathbb{C}^{r}$. We now give a precise proof of this statement.

\begin{Lemma}\label{LineBundleXnp} Let $(X,Y,\Delta,D,\tilde{D})$ be a quintuple satisfying $(\#)$ and $\pi: E \rightarrow Y$ a vector bundle of rank $r$. Then:
	\[i_{!}(E) \, \big\vert_{X_{D}^{n-p}} = i_{!}(Y \times \mathbb{C}^{r}) \, \big\vert_{X_{D}^{n-p}} \; .
\]
\end{Lemma}
\textbf{Proof:} referring to the notations in the proof of lemma \ref{Unitarity}, we have that:
	\[E \cdot \lambda_{\mathcal{N}} = i^{*} (\tilde{\pi}^{*})^{-1}(E \boxtimes \lambda_{\mathcal{N}}) = i^{*} (\tilde{\pi}^{*})^{-1} ( \pi_{1}^{*}E \otimes \pi_{2}^{*}\lambda_{\mathcal{N}} ) \; .
\]
Since $X^{n-p}_{D}$ intersects the tubular neighborhood in a finite number of cells, corresponding under $\varphi_{U}^{+}$ to a finite number of fibers of $\mathcal{N}$, it is sufficient to prove that, for any $y \in Y$, $(E \cdot \lambda_{\mathcal{N}}) \, \big\vert_{\mathcal{N}_{y}^{+}} = \lambda_{\mathcal{N}}^{\oplus r} \, \big\vert_{\mathcal{N}_{y}^{+}}$. First of all:
\begin{itemize}
	\item $i(\mathcal{N}_{y}^{+}) = (\{y\} \times \mathcal{N}_{y})^{+} \subset (\{y\} \times \mathcal{N})^{+}$;
	\item $E \cdot \lambda_{\mathcal{N}} \, \big\vert_{\mathcal{N}_{y}^{+}} = (i \vert_{\mathcal{N}_{y}^{+}})^{*} \bigl\{ \bigl[(\tilde{\pi}^{*})^{-1} (\pi_{1}^{*}E \otimes \pi_{2}^{*}\lambda_{\mathcal{N}})\bigr] \,\big\vert_{i(\mathcal{N}_{y}^{+})} \bigr\}$.
\end{itemize}
To obtain the bundle $\bigl[(\tilde{\pi}^{*})^{-1} (\pi_{1}^{*}E \otimes \pi_{2}^{*}\lambda_{\mathcal{N}} )\bigr] \,\big\vert_{i(\mathcal{N}_{y}^{+})}$, we can restrict $\tilde{\pi}$ to:
	\[A = \tilde{\pi}^{-1}[i(\mathcal{N}_{y}^{+})] = \tilde{\pi}^{-1}\bigl[(\{y\} \times \mathcal{N}_{y})^{+}\bigr] = \bigl(\{y\} \times \mathcal{N}_{y}^{+}\bigr) \,\cup\, \bigl(Y \times \{\infty\} \bigr) \,\cup\, \bigl(\{\infty\} \times \mathcal{N}^{+}\bigr)
\]
and consider $(\tilde{\pi} \, {\vert_{A}}^{*})^{-1} \bigl[( \pi_{1}^{*}E \otimes \pi_{2}^{*}\lambda_{\mathcal{N}} ) \,\big\vert_{A} \bigr]$. Moreover:
\begin{itemize}
	\item $(\pi_{1}^{*}E \otimes \pi_{2}^{*}\lambda_{\mathcal{N}}) \,\big\vert_{\{y\} \times \mathcal{N}_{y}^{+}} = (\mathbb{C}^{r} \otimes \pi_{2}^{*}\lambda_{\mathcal{N}} ) \,\big\vert_{\{y\} \times \mathcal{N}_{y}^{+}} \simeq \lambda_{\mathcal{N}}^{\oplus r} \,\big\vert_{\mathcal{N}_{y}^{+}}$;
	\item $(\pi_{1}^{*}E \otimes \pi_{2}^{*}\lambda_{\mathcal{N}}) \,\big\vert_{Y \times \{\infty\}} = (\pi_{1}^{*}E \otimes 0) \,\big\vert_{Y \times \{\infty\}} = 0$;
	\item $(\pi_{1}^{*}E \otimes \pi_{2}^{*}\lambda_{\mathcal{N}}) \,\big\vert_{\{\infty\} \times \mathcal{N}^{+}} = (0 \otimes \pi_{2}^{*}\lambda_{\mathcal{N}}) \,\big\vert_{\{\infty\} \times \mathcal{N}^{+}} = 0$.
\end{itemize}
Hence, since the three components of $A$ intersect each other at most at one point, by lemma \ref{OnePointUnion} we obtain:
	\[(\pi_{1}^{*}E \otimes \pi_{2}^{*}\lambda_{\mathcal{N}}) \,\big\vert_{A} = \bigl( \pi_{1}^{*}(Y \times \mathbb{C}^{r}) \otimes \pi_{2}^{*}\lambda_{\mathcal{N}} \bigr) \,\big\vert_{A} \; .
\]
$\square$

\paragraph{Remark:} In the statement of theorem \ref{SecondTheorem} (and of its generalization to any vector bundle) it was necessary to explicitly introduce a triangulation $\Delta$ on $X$, since the first step of the spectral sequence consists of simplicial cochains, which by definition depend on the simplicial structure chosen. Anyway, the groups $E^{p, \,0}_{r}$ for $r \geq 2$ and the filtration $\Ker( K(X) \rightarrow K(X^{n-p}))$ of $K(X)$ do not depend on the particular simplicial structure chosen \cite{AH}, thus, if we start from the cohomology class $\PD_{X}[Y]$ at the second step of the spectral sequence (which is the D-brane charge density with respect to the cohomological classification) we can drop the dependence on $\Delta$, $D$ and $\tilde{D}$. Therefore the choice of the triangulation has no effect on the physical classification of D-brane charges.

\subsection{Odd case}

We now consider the case of $n - p$ odd (for $n$ the dimension of $X$ and $p$ the dimension of $Y$), corresponding by a physical point of view to type IIA superstring theory. In this case the Gysin map takes value in $K^{1}(X)$, which is isomorphic to $K(\hat{S}^{1}X)$, for $\hat{S}^{1}X$ the unreduced suspension of $X$ defined as:
	\[\hat{S}^{1}X = (X \times [-1,1]) / (X \times \{-1\}, X \times \{1\})
\]
i.e.\ as a double cone built on $X$. We thus consider the natural embedding $i^{1}: Y \rightarrow \hat{S}^{1}X$ and the corresponding Gysin map:
	\[(i^{1})_{!}: K(Y) \rightarrow K(\hat{S}^{1}X) \simeq K^{1}(X) \; .
\]
Let $U$ be a tubular neighborhood of $Y$ in $X$, and let $U^{1} \subset \hat{S}^{1}X$ be the tubular neighborhood of $Y$ in $\hat{S}^{1}X$ defined removing the vertices of the double cone to $\hat{S}^{1}U$. We have that $\overline{\hat{S}^{1}(X^{n-p}_{D}\vert_{U})} \subset \overline{U^{1}}$ and $\hat{S}^{1}(X^{n-p-1}_{D}\vert_{\partial U}) \subset \partial U^{1}$, where $\partial U^{1}$ containes also the vertices of the double cone. In this way we can riformulate the previous results in the odd case, considering $\hat{S}^{1}(X^{n-p}_{D})$ and $\hat{S}^{1}(X^{n-p-1}_{D})$ rather than $X^{n-p}_{D}$ and $X^{n-p-1}_{D}$.

\paragraph{}We consider quintuples $(X, Y, \Delta, D, \tilde{D})$ safisfying the following condition:
\begin{itemize}
	\item[$(\#^{1})$] $X$ is an $n$-dimensional compact manifold and $Y \subset X$ a $p$-dimensional embedded compact submanifold, such that $n - p$ is \emph{odd} and $\mathcal{N}(Y)$ is spin$^{c}$. Moreover, $\Delta$, $D$ and $\tilde{D}$ are defined as in theorem \ref{Triangulation}.
\end{itemize}

\paragraph{}We now reformulate the same theorems stated for the even case, which can be proved in the same way. We remark that $\mathcal{N}_{\hat{S}^{1}X}Y$ is spin$^{c}$ if and only $\mathcal{N}_{X}Y$ is, since $\mathcal{N}_{\hat{S}^{1}X}Y = \mathcal{N}_{X}Y \oplus 1$ so that, by the axioms of characteristic classes \cite{MS}, $W_{3}$ is the same.

\begin{Lemma}\label{TrivialityXnp1Odd} Let $(X, Y, \Delta, D, \tilde{D})$ be a quintuple satisfying $(\#^{1})$ and $\alpha \in K(Y)$. Then:
\begin{itemize}
	\item there exists a neighborhood $V$ of $\hat{S}^{1}X \setminus U^{1}$ such that $i^{1}_{!}(\alpha) \, \big\vert_{V} = 0$;
	\item in particular, $i^{1}_{!}(\alpha) \, \big\vert_{\hat{S}^{1}(X^{n-p-1}_{D})} = 0$.
\end{itemize}
$\square$
\end{Lemma}

\begin{Theorem}\label{FirstTheoremOdd} Let $(X, Y, \Delta, D, \tilde{D})$ be a quintuple satisfying $(\#^{1})$ and $\Phi^{n-p}_{D}: C^{n-p}(X, \mathbb{Z}) \rightarrow K^{1}(X_{D}^{n-p}, X_{D}^{n-p-1}) \simeq K(\hat{S}^{1}(X_{D}^{n-p}), \hat{S}^{1}(X_{D}^{n-p-1}))$ be the isomorphism stated in theorem \ref{KTheoryCohomology}. Let:
	\[\pi^{n-p,\,n-p-1}: \hat{S}^{1}(X_{D}^{n-p}) \longrightarrow \hat{S}^{1}(X_{D}^{n-p}) / \hat{S}^{1}(X_{D}^{n-p-1})
\]
be the projection and $\PD_{\Delta}Y$ be the representative of $\PD_{X}[Y]$ (for $[Y]$ the homology class of $Y$) given by the sum of the cells dual to the $p$-cells of $\Delta$ covering $Y$. Then:
	\[i^{1}_{!}\,(Y \times \mathbb{C}) \, \big\vert_{\hat{S}^{1}(X_{D}^{n-p})} = (\pi^{n-p,\,n-p-1})^{*}( \Phi_{D}^{n-p}(\PD_{\Delta}Y)) \; .
\]
$\square$
\end{Theorem}

\begin{Theorem}\label{SecondTheoremOdd} Let $(X,Y,\Delta,D,\tilde{D})$ be a quintuple satisfying $(\#^{1})$ and $\Phi^{n-p}_{D}: C^{n-p}(X, \mathbb{Z}) \rightarrow K^{1}(X_{D}^{n-p}, X_{D}^{n-p-1})$ be the isomorphism stated in theorem \ref{KTheoryCohomology}. Let us suppose that $\PD_{\Delta}Y$ is contained in the kernel of all the boundaries $d^{n-p, \,0}_{r}$ for $r \geq 1$. Then it defines a class:
	\[\{\Phi^{n-p}_{D}(\PD_{\Delta}Y)\}_{E^{n-p, \,0}_{\infty}} \in E^{n-p,\,0}_{\infty} \simeq \frac{\, \Ker ( K^{1}(X) \longrightarrow K^{1}(X^{n-p-1}) ) \,} {\Ker ( K^{1}(X) \longrightarrow K^{1}(X^{n-p}) )} \; .
\]
The following equality holds:
	\[\{\Phi^{n-p}_{D}(\PD_{\Delta}Y)\}_{E^{n-p, \,0}_{\infty}} = [(i^{1})_{!}(Y \times \mathbb{C})] \; .
\]
$\square$
\end{Theorem}

\subsection{The rational case}

\subsubsection{Even case}

We now analyze the case of rational coefficients. We define:
	\[K_{\mathbb{Q}}(X) := K(X) \otimes_{\mathbb{Z}} \mathbb{Q} \; .
\]
We can thus classify the D-brane charge density at rational level as $i_{!}(E) \otimes_{\mathbb{Z}} \mathbb{Q}$. The Chern character provides an isomorphism $\ch: K_{\mathbb{Q}}(X) \rightarrow H^{\ev}(X, \mathbb{Q})$. Since the square root of $\hat{A}(TX)$ is a polyform starting with 1, it also defines an isomorphism, so that the composition:
	\[\begin{split}
	\widehat{\ch}: \,&K_{\mathbb{Q}}(X) \longrightarrow H^{\ev}(X, \mathbb{Q})\\
	&\widehat{\ch}(\alpha) = \ch(\alpha) \wedge \sqrt{\hat{A}(TX)}
\end{split}\]
remains an isomorphism. Thus, the classifications with rational K-theory and rational cohomology are completely equivalent.

We can also define the rational Atiyah-Hirzebruch spectral sequence $Q^{2k, \,0}_{r}(X) := E^{2k, \,0}_{r}(X)$ $\otimes_{\mathbb{Z}} \mathbb{Q}$. Such a sequence collapses at the second step \cite{AH}, i.e.\ at the level of cohomology: thus $Q^{2k, \,0}_{\infty}(X) \simeq Q^{2k, \,0}_{2}(X)$. An explicit isomorphism is given by the appropriate component of the Chern character:
	\[\ch_{\frac{n-p}{2}}: \frac{\, \Ker \bigl( K_{\mathbb{Q}}(X) \longrightarrow K_{\mathbb{Q}}(X^{n-p-1}) \bigr) \,} {\Ker \bigl( K_{\mathbb{Q}}(X) \longrightarrow K_{\mathbb{Q}}(X^{n-p}) \bigr)} \longrightarrow H^{n-p}(X, \mathbb{Q}) \; .
\]
This map is well-defined since, for a bundle which is trivial on the $(n-p)$-skeleton, the Chern characters of degree less or equal to $\frac{n-p}{2}$ are zero \cite{AH} (in particular $\ch_{\frac{n-p}{2}} = \widehat{\ch}_{\frac{n-p}{2}}$ for a bundle which is trivial on the $(n-p-1)$-skeleton). Moreover, since $Q^{2k,0}_{\infty}$ has no torsion:
	\[K_{\mathbb{Q}}(X) = \bigoplus_{2k}Q^{2k,0}_{\infty}
\]
and an isomorphism can be obtained splitting $\alpha \in K_{\mathbb{Q}}(X)$ as $\alpha = \sum_{2k}\alpha_{2k}$ where $\ch(\alpha_{2k}) = \ch_{k}(\alpha)$.

\subsubsection{Odd case}

In this case, we have the isomorphism $\ch: K^{1}_{\mathbb{Q}}(X) \rightarrow H^{\odd}(X, \mathbb{Q})$. Moreover, $H^{\odd}(X, \mathbb{Q}) \simeq H^{\ev}(\hat{S}^{1}X, \mathbb{Q})$. Hence we have the correspondence among:
\begin{itemize}
	\item $i^{1}_{!}(E) \in K^{1}_{\mathbb{Q}}(X)$;
	\item $\widehat{\ch}(i^{1}_{!}E) \in H^{\ev}(\hat{S}^{1}X, \mathbb{Q}) \simeq H^{\odd}(X, \mathbb{Q})$;
	\item $\oplus_{2k} \, \bigl[(i^{1}_{k})_{!}(Y_{k} \times \mathbb{C}^{q_{k}})\bigr]_{Q^{2k+1,0}_{\infty}}\,$.
\end{itemize}

\section{Conclusions and future perspectives}

To summarize, we have considered the classifications of D-brane charges in a compact euclidean space-time $S$ shown in table \ref{fig:Classifications2}.
\begin{table*}[h!]
	\centering
		\begin{tabular}{|l|l|l|}
			\hline & & \\ & Integer & Rational \\ & & \\ \hline
			& & \\ Cohomology & $\PD_{S}[q \cdot WY_{p}] \in H^{9-p}(S, \mathbb{Z})$ & $i_{\#}\bigl(\ch(E) \wedge G(WY_{p})\bigr) \in H^{\ev}(S, \mathbb{Q})$ \\ & & \\ \hline
			& & \\ K-theory (Gysin map) & $i_{!}(E) \in K(S)$ & $i_{!}(E) \in K_{\mathbb{Q}}(S)$ \\ & & \\ \hline
			& & \\ K-theory (AHSS) & $\{\PD_{S}[q\cdot WY_{p}]\} \in E^{9-p,\,0}_{\infty}(S)$ & $\{i_{\#}(\ch(E) \wedge G(WY_{p}))\} \in Q^{\ev,\,0}_{\infty}(S)$ \\ & & \\ \hline
		\end{tabular}
\caption{Classifications}\label{fig:Classifications2}
\end{table*}
We can now explain the relations between them. We already saw the complete equivalence of the three rational classifications, due to the isomorphisms $H^{*}(S, \mathbb{Q}) \simeq K^{*}_{\mathbb{Q}}(S) \simeq \bigoplus_{k}Q^{k,0}_{\infty}$, which split into even and odd parts. For the integral classifications, the three approaches are not equivalent, and our aim for this paper was to clarify their relationships. The cohomological and AHSS approaches have a clear link as one can see in the table, but they do not take into account the gauge and gravitational couplings. Since we have seen the link between Gysin map and Atiyah-Hirzebruch spectral sequence, we can also link the two corresponding approaches. We have proved that, for a world-volume $WY_{p}$ with gauge bundle $E$ of rank $q$, $i_{!}(E) \in \Ker (K^{9-p}(S) \rightarrow K^{9-p}(S^{8-p}))$ and that:
	\[\{\PD_{S}[q \cdot WY_{p}]\}_{E^{9-p, \,0}_{\infty}} = [i_{!}(E)] \; .
\]
Thus, we can use AHSS to detect possible anomalies, then we can use the Gysin map to get the charge of a non-anomalous brane: such a charge belongs to the equivalence class reached by AHSS, so that the Gysin map gives richer information. Some comments are in order. One could ask why the additional information provided by the Gysin map has to be considered: in fact, we have proven that it concerns the choice of a representative of the class, while, discussing AHSS in chapter 2, we have seen that one of its advantages is that it quotients out unstable configurations. It seems that such additional information keeps into account only instabilities. Actually, this is not the case. Let us consider a couple $(WY_{p}, i_{!}(E))$ made by a D-brane world-volume and its charge with respect to the Gysin map approach. The charge does not provide complete information about the world-volume, since $i_{!}E$ is a class in the whole space-time, exactly as the charge $q$ of a particle does not provide information about its trajectory. This is true also for the cohomological and AHSS classifications: two homologous world-volumes are not the same trajectory. If we consider two couples $(WY_{p}, i_{!}(E))$ and $(WY_{p}, i_{!}(F))$, we know that $[i_{!}(E) - i_{!}(F)]_{E^{9-p, \,0}_{\infty}} = 0$, which means that $i_{!}(E) - i_{!}(F)$ lies in the image of some boundaries of AHSS. Let us suppose that it lies in the image of $d_{3}$. This means that there exists an unstable world-volume $WU_{p}$ with a gauge bundle, e.g.\ the trivial one, such that $i_{!}(WU_{p} \times \mathbb{C}) = i_{!}(E) - i_{!}(F)$, but the two terms of the latter equality concern different world-volumes with the same zero charge: in fact, $WU_{p}$ has charge $0$ because it lies in the image of $d_{3}$, while $i_{!}(E-F)$ has charge $0$ since, being $\rk(E - F) = 0$, it is a representative of the class reached starting from $0\cdot WY_{p}$. Anyway, the wolrd-volume $WY_{p}$ is not anomalous in general and the fact that the gauge bundle on it is $E$ or $F$ is a meaningful information. Actually the information contained in $i_{!}(E-F)$ is partially contained in the charges of the sub-branes of $WY_{p}$. Thus, we can apply AHSS to the world-volume of the D-brane, then, if it corresponds to the trivial class we consider it as an unstable one, otherwise we can consider each representative of the class as an additional meaningful information.

\paragraph{}Possibile generalizations of this work are the following:
\begin{itemize}
	\item admitting the presence of the $B$-field compatibly with Freed-Witten anomaly, considering also twisted K-theory and the corresponding twisted AHSS;
	\item considering the case of non-compact space-time and world-volumes, using the appropriate form of AHSS;
	\item studying branes with singularities, using the appropriate form of the Gysin map.
\end{itemize}

\section*{Acknowledgements}

We would like to thank Loriano Bonora for the helpfulness he always showed since we started to work with him. We are also really grateful to Jarah Evslin for many suggestions and for having pointed out many subtleties. We also thank Giulio Bonelli and Ugo Bruzzo for useful discussions.

\end{document}